\documentclass[11pt]{article}

\usepackage[a4paper,margin=1in]{geometry}
\usepackage{amsmath,amssymb,amsthm}
\usepackage{caption}
\usepackage{subcaption}
\usepackage{tikz,xcolor,hyperref,color}

\usepackage{babel}
\usepackage{caption}
\usepackage{subcaption}
\usepackage[page,title]{appendix}
\usepackage{tikz,xcolor,hyperref,color}

\newcommand{\etal}{\textit{et al.} }
\newcommand{\popu}{_\mathrm{pop}}
\newcommand{\obs}{_\mathrm{obs}}

\renewcommand{\v}[1]{\ensuremath{\mathbf{#1}}} % for vectors
\newcommand{\gv}[1]{\ensuremath{\mbox{\boldmath$ #1 $}}} % for vectors of Greek letters
\newcommand{\mat}[1]{\mathrm{\mathbf{#1}}} % for matrices in Roman font
\newcommand{\grad}[1]{\gv{\nabla} #1} % for gradient
\newcommand{\abs}[1]{\left| #1 \right|} % for absolute value
\renewcommand{\d}[2]{\frac{d #1}{d #2}} % for derivatives

\newcommand{\edit}[1]{#1}

\newcommand{\quotes}[1]{``#1''}

\DeclareMathOperator{\logit}{logit}

\providecommand{\keywords}[1]{\textbf{Keywords: } #1}

\title{Exploring natural variation in tendon constitutive parameters via Bayesian data selection and mixed effects models}

\author{
James Casey$^{1}$,
Jessica Forsyth$^{1,2}$,
Timothy Waite$^{1}$,
Simon Cotter$^{1}$\thanks{Corresponding author: \texttt{simon.cotter@manchester.ac.uk}},
and Tom Shearer$^{1}$
}

\date{
$^{1}$ University of Manchester, Department of Mathematics, Manchester, M13 9PL, UK \\
$^{2}$ University of Sheffield, School of Medicine and Population Health, S10 2TN, UK
}

\begin{document}
\maketitle

\begin{abstract}
Combining microstructural mechanical models with experimental data enhances our understanding of the mechanics of soft tissue, such as tendons. In previous work, a Bayesian framework was used to infer constitutive parameters from uniaxial stress-strain experiments on horse tendons, specifically the superficial digital flexor tendon (SDFT) and common digital extensor tendon (CDET), on a per-experiment basis. Here, we extend this analysis to investigate the natural variation of these parameters across a population of horses. Using a Bayesian mixed effects model, we infer population distributions of these parameters. Given that the chosen hyperelastic model does not account for tendon damage, careful data selection is necessary. Avoiding ad hoc methods, we introduce a hierarchical Bayesian data selection method. This two-stage approach selects data per experiment, and integrates data weightings into the Bayesian mixed effects model. Our results indicate that the CDET is stiffer than the SDFT, likely due to a higher collagen volume fraction. The modes of the parameter distributions yield estimates of the product of the collagen volume fraction and Young’s modulus as 811.5 MPa for the SDFT and 1430.2 MPa for the CDET. This suggests that positional tendons have stiffer collagen fibrils and/or higher collagen volume density than energy-storing tendons.
\end{abstract}

\keywords{Tendon, data selection, mixed effects, Bayesian inference, nonlinear elastic}

\section{Introduction}
\label{sctn:introduction}
Accurate description of the nonlinear mechanical behaviour of collagen is important for understanding and predicting the properties of a wide range of soft tissues, including arterial walls \cite{holzapfel2010,Mishani2021}, skin \cite{Limbert2019,PISSARENKO2020208}, and tendons \cite{shearer2015new,shearer2015newhelical,shearer2020recruitment}. This knowledge is vital for designing artificial tissues for grafts and surgical interventions \cite{Theodossiou2019}. In this paper, we focus on a model for tendons; however, the underlying approach could be extended to other soft tissues.

Bayesian inverse methods are increasingly used in biomechanics to estimate soft tissue parameters while quantifying uncertainty. Such approaches have been applied to infer viscoelastic properties from acoustic radiation force imaging \cite{zhao2016bayesian}, and to calibrate hyper-viscoelastic models of brain tissue under varying experimental protocols \cite{teferra2019bayesian}. A Bayesian framework has also been used for model selection and sensitivity analysis in studies of the knee meniscus \cite{elmukashfi2022model}. These examples illustrate the value of Bayesian inference for parameter identification in complex, heterogeneous tissues.

\begin{figure}[h]
    \centering
    \includegraphics[width=0.55\linewidth]{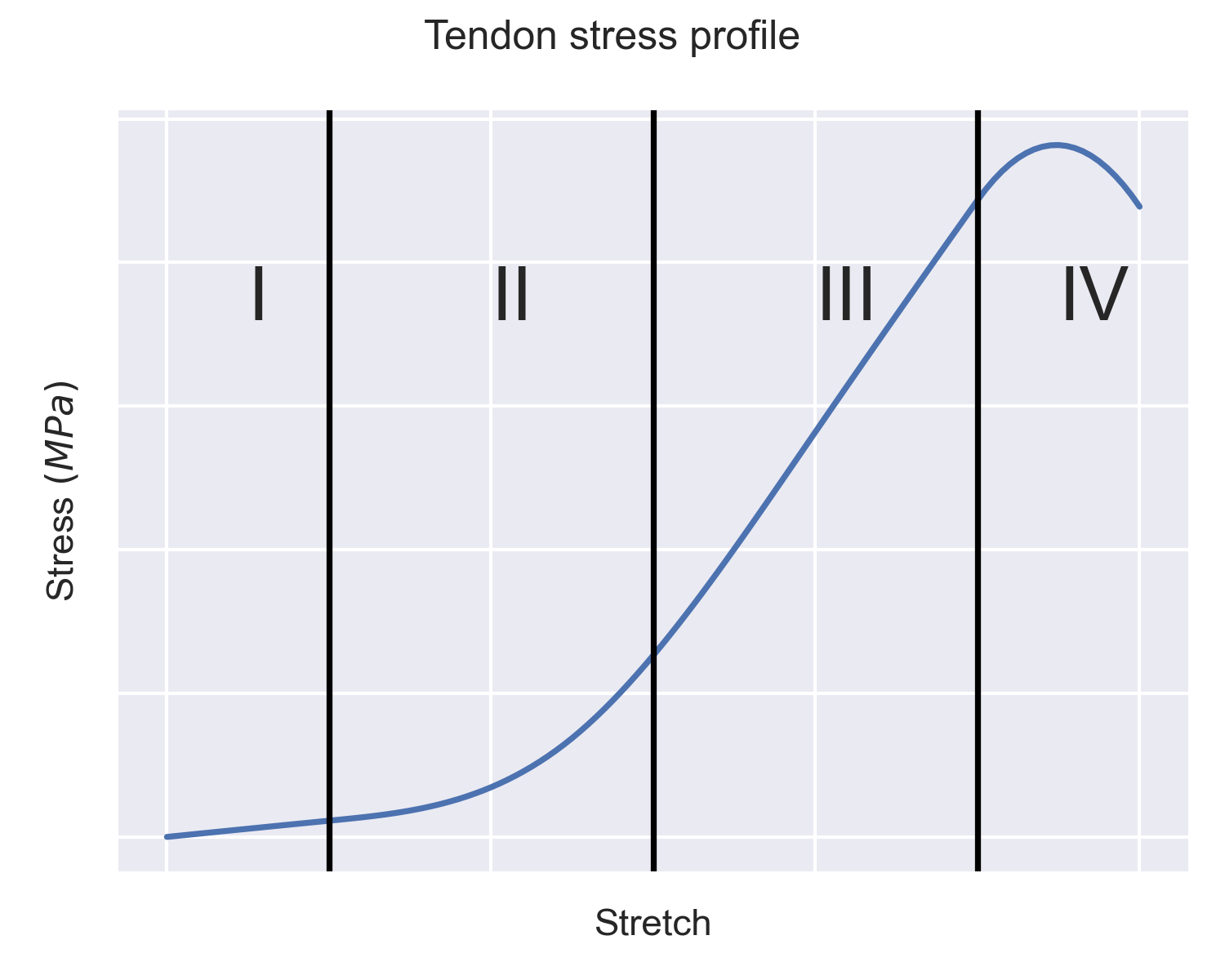}
    \caption{Idealised stress profile of a tendon. The four regions are: I) Toe, II) Heel, III) Linear, IV) Damage. In reality, damage may begin occurring in region III, or even region II, due to the shorter fibrils breaking.}
    \label{fig:stress_profile}
\end{figure}

\begin{figure}
    \centering
\includegraphics[width=0.9\textwidth]{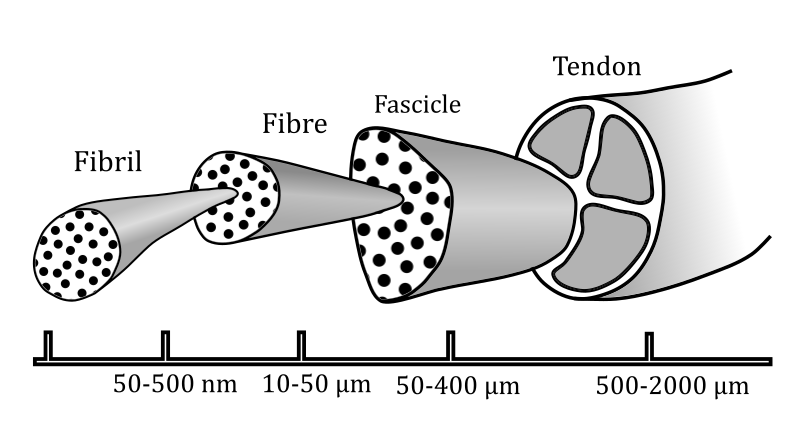}
    \caption{Microstructure of a typical tendon. Bundles of fibrils form fibres, which are embedded within fascicles, which make up the tendon. The NCM is the non-collagenous portion of the tendon. Adapted from Jull \etal \cite{tendonImage}.
    }
    \label{fig:tendon_structure}
\end{figure}

Tendons are fibrous tissues that connect, and transfer forces between, muscles and bones \cite{shearer2015new,WANG20061563}. Their complex microstructure gives rise to anisotropy and non-linear stress-strain profiles \cite{HAKEN2000195,BOL20151092,Hoffmeister19963933}. An example of a typical stress-strain profile for a tendon undergoing a uniaxial stretch along its longitudinal axis is given in Figure \ref{fig:stress_profile}. Tendons consist of collagen fibrils of varying lengths that assume a crimped waveform within collagen fibres. The fibres are embedded within a non-collagenous matrix (NCM), and reinforce the tendon along a preferred axis \cite{shearer2015new,haughton}, which causes anisotropy. A schematic representation of a tendon's microstructure is shown in Figure \ref{fig:tendon_structure}. As the tendon is stretched, the fibrils begin to straighten and contribute to the stress response. Due to the fibrils having varying lengths, their recruitment is gradual, which results in the nonlinearity of the stress profile.

The characteristic shape of the stress-strain graph has four regions: I) Toe II) Heel, III) Linear, and IV) Damage, with each region corresponding to a different physical phenomenon. The toe region corresponds to the fibrils being slack, resulting in a stress profile that is approximately linear, as governed by the NCM. In the heel region, fibrils are gradually becoming taut and contributing to the stress response, resulting in a non-linear profile. In the linear region, all of the fibres have been recruited and the stress profile is primarily governed by the fibril stress response. In the final region, failure occurs, the tendon is damaged, fibrils begin to break gradually and the tendon no longer deforms elastically.

In Figure \ref{fig:stress_profile}, the stress response is idealised. In reality, fibrils may begin breaking in earlier regions (III or even II), resulting in data in region III being recorded which is not consistent with models which neglect damage. Typically, the data is manually trimmed to a specified limit \cite{ShearerHazelScreen2017}. This approach tends to include a large portion of the linear region (region III), and fitting models to this data can lead to inaccurate estimates of various parameters in the presence of early fibril damage. For this reason, when fitting models to tendon stress-strain data, it is beneficial to use a more sophisticated data selection method to include as much data as possible that is valid under the assumptions of the mechanical model used, whilst down-weighting the contribution of data points that do not satisfy the model's assumptions \cite{cotter2022hierarchical}.

Models for the stress response of tendons often describe the deformation in terms of the physical parameters of their constituents, such as the Young's modulus and shear modulus. These models are known as microstructural models, and have been used by several authors for various soft tissues \cite{shearer2015new, CIARLETTA20062034}. 

One of the advantages of microstructural modelling is that the model parameters have a direct physical meaning, and can be measured experimentally. However, a problem that arises is that a wide range of values is often reported for the same parameter. The Young's modulus of Type 1 collagen fibrils, for example, has been reported to be as low as 32 MPa \cite{GRAHAM2004335}, and as high as 2.8 GPa \cite{svensson}. This spread of reported values could be due to several factors, such as differences in the ages of the samples, inter-species variation, model misspecification within the inference, or even that the parameters may be poorly identifiable from the data.
Due to this, it is often not entirely clear what the credible range for a given parameter value is, nor which values are most likely. Quantification of this spread is key to understanding the natural behaviour of soft tissues and their constituents. In this paper, we tackle the problem of \textit{intra}-species variation. That is, within a given species, individuals can yield varying values for constitutive parameters due to naturally occurring heterogeneity within the population. For context, the individuals in this study are horse tendons, and the species are the SDFT and CDET types.

To tackle the uncertainty in the parameter values, we employ a Bayesian approach. This provides a rigorous framework with which to combine prior knowledge, mechanical models, and data. Typically, inference is performed on data from a single individual from the population at a time, as in previous work~\cite{haughton}. In this paper, we instead use data from multiple individuals to infer the population-level variation in the constitutive parameters of a microstructural, nonlinear elastic tendon model, through a statistical mixed effects model.

Since fibril damage occurs at different points for different individuals, we are prompted to use a more sophisticated data selection technique than uniform trimming of all data sets (for example, after a certain strain has been reached \cite{ShearerHazelScreen2017}). We implement a hierarchical Bayesian data selection technique, which allows us to infer which data are consistent with the chosen microstructural model \cite{cotter2022hierarchical}. We use a two-stage approach. The first stage involves data selection on a per-individual basis, leading to a re-weighting of each of the observations in the likelihood, along with an automatically chosen truncation of the data. The second stage uses the data from all individuals (the population), with re-weighted observations and truncations, in the population-level inference, which computes the population posterior distribution for the microstructural model parameters. The reason for performing the inference in two stages is that it would be prohibitively costly to simultaneously perform data selection and calculation of mixed effects posteriors. Moreover, the trimming of data with low weightings which have large discrepancies with the chosen mechanistic model improves overall model fit and reduces the bias in the overall inference \cite{cotter2022hierarchical}.

We employ a mixed effects model in order to infer from multiple individuals simultaneously. This allows us to pool information and build knowledge of population-level distributions of the parameters in addition to the parameter distributions of each individual. Mixed effects models have previously been used to study soft tissue mechanical parameters \cite{Wang8626521,WANG20171}, as has the Bayesian framework \cite{WANG20061563,haughton,WANG2023106070}; however, to the authors' knowledge, there are no examples of a Bayesian mixed effects model being used for soft tissue inference in the literature. Y.N. Wang \textit{et al.} \cite{WANG20171} used a linear mixed effects model to infer model parameters without any Bayesian modelling, whilst P. Wang \textit{et al.} \cite{WANG2023106070} used Bayesian modelling to infer model parameters without incorporating mixed effects. Bayesian mixed effects models \textit{have} been used across many other areas of study, however, such as in recommender systems \cite{condliff1999bayesian}, neurology \cite{ziegler2015estimating}, and pharmacokinetics \cite{patoux2001comparison}. 

The structure of the paper is as follows. Section \ref{sctn:tendon_model} introduces the microstructural model. To quantify the variability and uncertainty in the model parameters, we describe the Bayesian approach in Section \ref{sctn:bayes}. The data we use in this study is outlined in Section \ref{sctn:data}. To calculate the resulting posterior distributions, we use Markov chain Monte Carlo (MCMC) sampling. We ensure that we are inferring model parameters using data where the microstructural model is valid, for example before a considerable amount of damage has occurred, by implementing a Bayesian data selection technique, described in Section \ref{sctn:fidel_params}. To model the heterogeneity of the physical parameters across individuals in the population, a mixed effects model is used, which we discuss in Section \ref{sctn:me_model}. We discuss our choice of priors for the parameters in Section \ref{sctn:prior_selection}, and in Section \ref{sctn:sampling_with_stan}, we discuss the implementation of the Hamiltonian Monte Carlo (HMC) sampler using the Stan package \cite{Stan}, and also discuss sampler parameter tuning in order to ensure convergence of the Markov chains. The results of our modelling approach are given in Section \ref{sctn:results}. We finish with a discussion in Section \ref{sctn:discussion}.

\section{Methods}
\label{sctn:methods}
In this paper, we are considering the inverse problem of inferring model parameters from observational data. There are many approaches to tackling inverse problems, including maximum likelihood estimation and Bayesian maximum \textit{a posteriori} estimation. In many circumstances, quantifying the uncertainty in parameter estimates is of interest, for example by estimating credible ranges for the model parameters. One approach to achieving this is to use a Bayesian framework to determine model parameter posterior distributions from the data. Typically, these distributions are not analytically computable and require approximation via numerical methods such as MCMC. While MCMC is the gold standard for Bayesian inverse problems, it incurs a high computational cost due to the large number of forward model evaluations required. However, if it is computationally feasible to implement for a given model, the samples produced can be used to provide detailed statistical information, including credible regions for the model parameters~\cite{neal2011}.

We also consider the problem of selecting data for inference where a subset of the data is known not to be consistent with the model. To do this, we again employ a Bayesian framework, this time to infer the values of parameters which convey how well the model fits each data point. These parameters can then be used to tune out data with mild discrepancies, or to trim data with moderate to severe discrepancies with the model.

\subsection{Microstructural Tendon Model}
\label{sctn:tendon_model}
To study the deformation of tendons, we use the microstructural model introduced by Haughton \etal \cite{haughton}. We assume the tendons under consideration can be modelled as circular cylinders, with fibres oriented along their long axes. Using cylindrical polar coordinates and denoting the deformed and undeformed configurations with lower- and upper-case letters respectively, the deformation can be described as $(r,\,\theta,\,z) = (\lambda^{-\frac{1}{2}}R,\,\Theta,\, \lambda Z)$, where $\lambda$ is the longitudinal stretch. The deformation gradient is defined as the gradient of the deformed position vector $\v{x}$ with respect to the undeformed coordinates, denoted
\begin{equation}
    \mat{F} = \grad \v{x},
\end{equation}
where $\grad$ is the gradient operator with respect to the undeformed coordinates. For the prescribed uniaxial deformation, it takes the form
\begin{equation}
    \mat{F} = \begin{pmatrix}
        \lambda^{-\frac{1}{2}} & 0 & 0\\
        0 & \lambda^{-\frac{1}{2}} & 0\\
        0 & 0 & \lambda
    \end{pmatrix}.
\end{equation}

From $\mat{F}$, we can calculate the left Cauchy-Green strain tensor $\mat{B} = \mat{F}\mat{F}^T$. Denoting the undeformed fibril orientation as $\v{M}$, we can calculate the deformed fibril orientation $\v{m} = \mat{F} \v{M}$ and define invariants of the deformation $I_1,\,I_2,\,I_3,\,I_4$, and $I_5$ as
\begin{align}
\begin{aligned}
    I_1 &= \mathrm{tr}(\mat{B}),\\
    I_2 &= \frac{1}{2} \left( \mathrm{tr}(\mat{B})^2 - \mathrm{tr}(\mat{B}^2) \right),\\
    I_3 &= \mathrm{det}(\mat{B}),\\
    I_4 &= \v{M} \cdot \mat{B}\v{M},\\
    I_5 &= \v{M} \cdot \mat{B}^2\v{M}.
\end{aligned}
\end{align}

We assume that tendons are incompressible, so that $I_3=1$, and follow the widely-used assumption that the deformation of the NCM can be described solely in terms of $I_1$, and that of the collagen fibrils in terms of $I_4$ \cite{shearer2015new,haughton,hgopaper}. Further, we assume that the strain energy function decomposes additively into contributions from the NCM and the fibrils, each scaled by their respective volume fractions. Denoting $\phi$ as the collagen volume fraction, these are $1 - \phi$ and $\phi$, respectively. It is assumed that the NCM is neo-Hookean with shear modulus $\mu$, whilst the fibrils are assumed to be Hookean, with Young's modulus $E$, after the macroscale stretch reaches a critical value $\lambda_C$, which is treated as a random variable (the value of $\lambda_C$ corresponds to the length of each fibril relative to the section of tendon in which it is embedded, with longer fibrils having higher values of $\lambda_C$). The critical stretch is assumed to have a triangular distribution, with lower limit $a$, upper limit $b$, and mode $c$.

From these assumptions, the resulting strain energy function \cite{haughton} is given by
\begin{align}
\begin{aligned}
    W(I_1,\,I_4) = (1 - \phi) \frac{\mu}{2} \left( I_1 - 3 \right) &+ \phi E\biggl( \frac{A(I_4,\, a,\,b,\,c)}{2} \log{I_4} + \left( B(I_4,\, a,\,b,\,c) - D(I_4,\, a,\,b,\,c)\right)\sqrt{I_4}\\
    &+ \frac{C(I_4,\, a,\,b,\,c)}{2}I_4 + \frac{D(I_4,\, a,\,b,\,c)}{2} \sqrt{I_4}\log{I_4} + G(I_4,\, a,\,b,\,c)\biggr),
\end{aligned}
\end{align}
where $A$, $B$, $C$, $D$ and $G$ are piecewise constant functions of $I_4$, which are defined in \cite{haughton}. Using this strain energy function, the only non-trivial entry of the engineering stress tensor, $N$, can be calculated in terms of the applied stretch $\lambda$ as
\begin{align}
\begin{aligned}
    N(\lambda,\,(1 - \phi)\mu,\,\phi E,\, a,\,b,\,c) = (1 - \phi)\mu\left(\lambda - \frac{1}{\lambda^2}\right) &+ \frac{\phi E}{\lambda} \biggl(A(\lambda^2,\, a,\,b,\,c) + B(\lambda^2,\, a,\,b,\,c)\lambda\\
      &+ C(\lambda^2,\, a,\,b,\,c)\lambda^2 + D(\lambda^2,\, a,\,b,\,c)\lambda\log{\lambda}\biggr).
\end{aligned}
\label{eq:haughton_stress}
\end{align}
A full derivation of this relationship is available in \cite{haughton}. For our statistical inference, we use the stress-stretch formula $N(\cdot)$ as the response function.

\subsubsection{The Linear Modulus}
\label{sctn:linear_modulus}
The linear modulus is the gradient of the stress in the linear region. Assuming a small contribution of the NCM to the stress response (typically $\frac{E}{\mu} \approx 10^6$ \cite{ShearerHazelScreen2017}), the linear region is dominated by the fibril stress. Under this assumption, the linear modulus, $LM$, can be calculated by Taylor series expanding the linear region of the fibril contribution to the engineering stress,
\begin{equation}
    N_\mathrm{linear}(\lambda) = \frac{\phi E}{\lambda} \left(-1 + 4 \lambda \frac{a \log\{\frac{a}{c}\} + b \log\{\frac{b}{c}\}}{(a - b)^2}\right),
\end{equation} 
about the stretch at which it is being calculated, $\bar{\lambda}$, say, and taking the coefficient of the linear term, which gives
\begin{equation}
    LM =\left.\d{N_\mathrm{linear}(\lambda)}{\lambda}\right|_{\lambda=\bar{\lambda}} = \frac{\phi E}{\bar{\lambda}^2}.
\end{equation}
We see that in the linear region of our model, the linear modulus is proportional to $\phi E$, but since $\bar{\lambda}>1$, it will always be less than $\phi E$. In Section \ref{sctn:me_model_results}, we will compare \edit{direct} estimates of the linear modulus with our inferred values of $\phi E$.

\subsection{Bayesian Inference}
\label{sctn:bayes}
We wish to infer the unknown model parameter values, given the data $\v{y}$, using the model $N(\cdot)$. To simplify the model, and to reduce the dimension of the target parameter space, we assume that the distribution of the critical stretch is symmetric, meaning that the parameter $c$ is completely determined by $a$ and $b$ as $c = \frac{a+b}{2}$,  and we therefore infer the parameters using the model $\mathcal{M}(\lambda,\, (1 - \phi)\mu,\, \phi E,\, a,\, b) = N(\lambda,\, (1 - \phi)\mu,\, \phi E,\, a,\, b, \frac{a + b}{2})$. We denote the model parameters of interest as $\gv{\theta} = [(1 - \phi)\mu,\,\phi E,\,a,\,b] \in \mathbb{R}^2_{\geq 0} \times \mathbb{R}^2_{\geq 1}$. Based on initial exploration, we concluded that the parameter $\phi$ was not identifiable from the data, and so we cannot infer $\phi$, $\mu$, and $E$ independently. Fortunately, measurements and estimations of $\phi$ can be made in a lab setting, however this is a destructive process and therefore separate representative samples must be destroyed in order to measure $\phi$ \cite{goh2018}.

In the Bayesian framework, our knowledge about the parameters $\gv{\theta}$, is captured via the posterior probability distribution, whose density $\pi(\gv{\theta} \,|\, \v{y};\, \mathcal{M})$ is obtained by combining prior knowledge with information from the observed data, $\v{y}$. Prior knowledge is encoded through the choice of an appropriate prior distribution, whose probability density we denote as $\pi_0(\gv{\theta})$. The contribution from the data is captured by the likelihood, denoted $\mathcal{L}(\v{y} \,|\, \gv{\theta};\, \mathcal{M})$. Bayes' theorem then allows us to express the posterior density as
\begin{equation}
    \pi(\gv{\theta} \,|\, \v{y};\, \mathcal{M}) \propto \mathcal{L}(\v{y}\,|\,\gv{\theta};\,\mathcal{M})\,\pi_0(\gv{\theta}),
    \label{eq:bayesProportionality}
\end{equation}
which yields an expression for the unknown density $\pi(\gv{\theta} \,|\, \v{y};\, \mathcal{M})$ in terms of the known densities $\mathcal{L}(\v{y}\,|\,\gv{\theta};\,\mathcal{M})$ and $\pi_0(\gv{\theta})$, up to a constant of proportionality. Since we only consider a single model in this study, henceforth, the dependence on the model is left implicit.

\subsection{Data}
\label{sctn:data}
\begin{figure}[h]
    \centering
    \includegraphics[width=\linewidth]{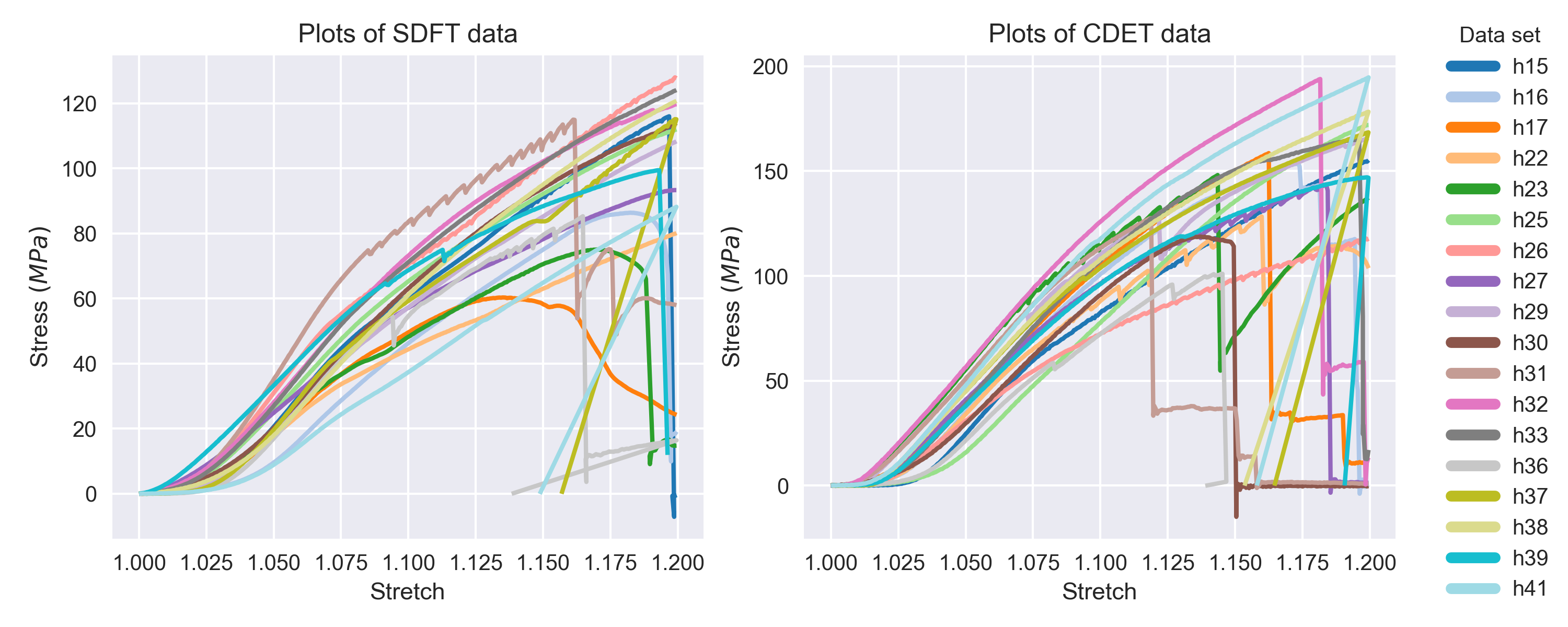}
    \caption{Plots of experimental tensile test data for the two tendon types: SDFT (left) and CDET (right). The data was manually trimmed to 20\% strain (equivalent to a stretch of 1.2). Each experiment is labelled \quotes{hXX}, where XX are numbers that refer to the label given to each horse by Screen \etal \cite{Thorpe2012}.}
    \label{fig:joint_data_plot}
\end{figure}

The experimental data that we will use to infer the constitutive parameters of our tendon model, which was collected by Thorpe \etal \cite{Thorpe2012}, consists of tensile tests applied to two types of tendons harvested from 18 different horses. The two types of tendons were the superficial digital flexor tendon (SDFT), and the common digital extensor tendon (CDET). SDFTs are energy-storing tendons, whereas CDETs are positional tendons \cite{Thorpe2012}. Figure \ref{fig:joint_data_plot} shows plots of all 18 tests for the two tendon types, manually trimmed to 20\% strain. It is clear that there is a common trend in the stress profiles of each experiment which indicates that it may be possible to use the same model of deformation to infer subject-specific model parameters for each individual tendon. Heterogeneity of the model parameters is present as can be seen by the data having varying gradients in the linear region (region III, governed by the product $\phi E$), and the onset of the heel region (region II, governed by the parameters of the triangular distribution $a$ and $b$). Since the data were collected by the same team using the same equipment, we assume that all experiments have the same observational noise.

Another feature of the data is heterogeneity in the suspected onset of damage between experiments, as indicated by the changing gradient, and differing failure stretches. For most experiments, the damage region appears to begin somewhere in the stretch range of 1.06 to 1.10, whereas a few experiments appear to have damage potentially occurring much earlier, at stretches of around 1.03. Clearly, trimming all of the experiments to a singular stretch value would not accurately remove the entire damage region for all tendons without also removing large portions of the elastic region for many of them. Using subjective methods to pick the point at which to trim the data for each experiment has the potential to lead to biased and/or more uncertain estimates of the model parameters, depending on whether the truncation is under or over-zealous. We are therefore motivated to explore a more sophisticated methodology to quantify where the data can and cannot be well represented by the chosen mechanistic model. To do so, in the next section, we adapt a hierarchical Bayesian data selection method, as laid out in \cite{cotter2022hierarchical}, to identify the regions of data where damage has occurred and the data-model discrepancy is high. These methods were first applied to unlabelled landmark matching of digital images, specifically matching biological cell clusters imaged via different modalities \cite{forsyth2023unlabelled}.

\subsection{Bayesian Data Selection}
\label{sctn:fidel_params}
In previous work by Haughton \etal \cite{haughton}, a Bayesian framework was used to infer the microstructural model parameters and their associated uncertainty. In an attempt to remove data where the fibrils had started to be damaged, the data was truncated at 10\% strain. In the entire population dataset shown in Figure \ref{fig:joint_data_plot}, however, it is evident that, in some tendons, damage initiates prior to 10\% strain, and thus an elastic model would not fit all of this data well, whereas in other cases, the elastic region appears to continue beyond 10\% strain and thus there would potentially be unused valid data if all curves were truncated at 10\%.

In this work, we aim to use the data from all 18 of the tensile experiments on two types of horse tendon, and we are therefore required to select data from a total of 36 experiments. To do this in an automated and objective manner, we employ hierarchical Bayesian data selection, which enables identification of regions where there is good agreement between the model and data, thereby eliminating the need for subjective manual truncation of the data prior to model fitting. The method involves the introduction of parameters which indicate the fidelity of the model for each data point, and thus greatly increases the dimensionality of the target distribution. Unfortunately, this means that it becomes computationally infeasible to simultaneously incorporate both Bayesian data selection and the mixed effects model, which we describe in Section \ref{sctn:me_model}. Instead, we propose a two-stage process in which we first conduct Bayesian data selection on a per-individual basis, identifying which data to trim, and re-weighting the remaining data in the likelihood. In the second stage, we use trimmed data, and the posterior means of the fidelity weights, within the mixed effects model.

\subsubsection{Definition of the likelihood}

Each experiment corresponds to a single tendon, and consists of a sequence of measurements of the stress, $y_j$, required to achieve stretch $\lambda_j$ ($j=1,\ldots,N)$, where $N \in \mathbb{N}$ is the number of observations in the data. We assume that each observation is subject to additive, zero mean, Gaussian, independent and identically distributed noise, that is
\begin{equation}
    y_j = \mathcal{M}(\lambda_j,\,\gv{\theta}) + \eta_j, \qquad \eta_j \sim \mathcal{N}(0, \sigma\obs^2)\,,
\end{equation}
and we let $\v{y} = [y_1, \ldots, y_N]^\top$. Therefore, the likelihood without data selection is given by
\begin{equation}
    \mathcal{L}(\v{y} \,|\, \gv{\theta},\,\sigma\obs^2) =  \prod_{j=1}^{N} \frac{1}{\sqrt{2\pi \sigma\obs^2}} \exp \left( -\frac{1}{2\sigma\obs^2}| y_j - \mathcal{M}(\lambda_j,\, \gv{\theta})|^2 \right)\,.
    \label{eq:likelihood_orig}
\end{equation}

We fix $\sigma\obs^2$ to a value that was estimated by fitting the model to very low strain data for all tendons. Then we introduce `fidelity' parameters for each data point, following the naming convention of previous works \cite{cotter2022hierarchical,forsyth2023unlabelled}, which indicate the model's ability to represent a single observation in the data. 
The fidelity parameters take values $\gv{\gamma} \in (0,1)^{N}$, where values close to 0 correspond to tuning out that data point's contribution to the likelihood and values close to 1 correspond to a standard contribution to the likelihood (as in the absence of data selection). This is implemented through a modification to the likelihood:
\begin{subequations}
\begin{align}
    \mathcal{L}(\v{y} \,|\, \gv{\theta},\, \sigma\obs^2,\, \gv{\gamma}) &= \prod_{j=1}^{N} \frac{1}{\sqrt{2\pi \sigma\obs^2 \gamma_j^{-1} }} \; \exp \left( -\frac{\gamma_j}{2\sigma\obs^2}\| y_j - \mathcal{M}(\lambda_j,\, \gv{\theta})\|^2 \right),\\
    &\propto  \; \prod_{j=1}^{N}  \frac{\sqrt{\gamma_j}}{\sigma\obs} \; \exp \left( -\frac{\gamma_j}{2\sigma\obs^2}\| y_j - \mathcal{M}(\lambda_j,\, \gv{\theta})\|^2 \right).
    \end{align}
\label{eq:fidlikelihood}
\end{subequations}
This approach is similar to that of power likelihoods in a generalised Bayesian framework \cite{holmes2017assigning}, except that instead of picking a single value for the exponent of the likelihood, we have the additional flexibility of having different values of the likelihood exponent for each individual observation in the data. Our aim in hierarchical Bayesian data selection is to infer the values of the $\gamma_j$.

\subsubsection{Data selection prior}
By invoking Bayes' rule, the posterior is given by: 
\begin{equation}
\pi(\gv{\theta}, \sigma\obs^2, \gv{\gamma}) \propto \pi_0(\gv{\theta}) \;  \pi_0(\gv{\gamma}) \;  \prod_{j=1}^{N}  \;  \frac{\sqrt{\gamma_j}}{\sqrt{2\pi \sigma\obs^2}}\; \; \exp \left( -\frac{\gamma_j}{2\sigma\obs^2}\| y_j - \mathcal{M}(\lambda_j,\, \gv{\theta})\|^2 \right), 
\label{eq:posterior}
\end{equation} 
where $\pi_0(\gv{\theta})$ is the prior for the microstructural parameters $\gv{\theta}$ as described in Section \ref{sctn:prior_selection}, and $\pi_0(\gv{\gamma})$ is the prior on the fidelity parameters.

The prior we define on $\gv{\gamma}$ must  be supported on $(0,1)^{N}$, encode some correlation between $\gamma_{j}$ and $\gamma_{j'}$ as a function of $|\lambda_j - \lambda_{j'}|$, dependent on a given length scale, and reflect our knowledge that there is unlikely to be damage occurring within the region of data with  $<$5\% strain. We adopt a logit Gaussian process prior for the fidelity field, which describes the prior fidelity at all possible points in the observation space~\cite{cotter2022hierarchical}. Since we are only interested in the values of the fidelity field at our observation points, this logit Gaussian process prior collapses down to a logit multivariate normal prior on $\gv{\gamma}$, where the logit function is applied element-wise.
Therefore, the prior on the transformed fidelity parameters, $\gv{\chi}_{\gv{\gamma}} = \logit(\gv{\gamma})$, is given by a multivariate normal distribution with mean $\gv{\mu}$ and covariance matrix $\mat{\Sigma}_{\gv{\gamma}}$.

One problem that can occur if naively applying Bayesian data selection is that, in some experiments, there is some early damage, and then a considerable amount of additional strain before further damage or failure occurs. In this scenario, without the application of an appropriate prior, the Bayesian data selection posterior may select the data following the initial damage, leading to low fidelity means for the early observations, which we know are unlikely to have been subject to damage. In order to mitigate this, we choose $\mu$ and $\sigma_\gamma$, which define the multivariate normal prior on the transformed fidelity parameters, such that the fidelity parameters have higher mean values for observation points with low strain. 

Therefore, we allow $\mu_\gamma(\lambda)$, as well as the standard deviation, $\sigma_\gamma(\lambda)$, to vary with strain. 
We relate the mean and standard deviations to strain via
\begin{subequations}
\begin{align}
    \mu_\gamma(\lambda) & = A_\mu + \frac{K_\mu-A_\mu}{1+\exp\left(-B(\lambda - \lambda_0)\right)},
    \label{eq:musigmoid}\\
    \sigma_\gamma(\lambda) & = A_\sigma + \frac{K_\sigma-A_\sigma}{1+\exp\left(-B(\lambda - \lambda_0)\right)}, 
    \label{eq:sigmasigmoid}
    \end{align}
\end{subequations}
where $A_{\mu}$, $A_{\sigma}$ and $K_{\mu}$, $K_{\sigma}$ are the left/right asymptotic values of $\mu$ or $\sigma$ respectively, $B$ is the rate of decrease/increase between the two asymptotes, and $\lambda_0$ is chosen as the strain value at which we believe that damage is likely to have started to have an effect. 

\begin{figure}
    \centering
\includegraphics[width=0.65\textwidth]{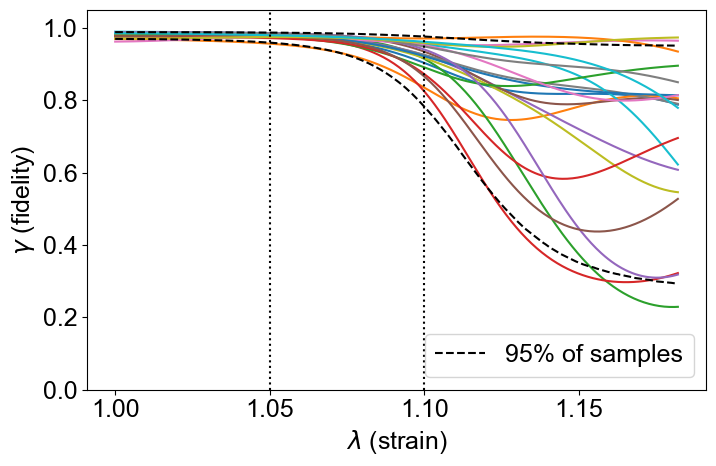}
    \caption{Twenty realisations of the logit-multivariate normal prior on the data fidelity parameters with respect to tendon strain. Dashed line represents the 95\% confidence interval of the prior samples.}
    \label{fig:squidprior}
\end{figure}

To construct the covariance matrix, $\Sigma_\gamma$, which is $N \times N$, and encodes the correlation between fidelity parameters, we use a vertical scaling of a squared exponential kernel such that 
\begin{equation}
    \Sigma_\gamma[j,j'] = \sigma_\gamma(\lambda_j) \cdot \sigma_\gamma(\lambda_{j'}) \cdot \exp \left(- \frac{|\lambda_j - \lambda_{j'} |^2_2 }{2 l^2} \right),
\end{equation}
where $l$ is the lengthscale of the kernel and describes the length of correlation between fidelity parameters, and $\sigma_\gamma(\cdot)$ is the standard deviation at a given strain value according to Equation~\eqref{eq:sigmasigmoid}. We choose $A_\mu=4, K_\mu=1, A_\sigma=0.25, K_\sigma=1, B=50, \lambda_0=1.1, l =0.05$ to ensure we have high prior mean and low variance for strains less than 5\% where we expect significant damage to the tendon to be rare, and high variance (and therefore lower mean) for strains above 10\% where damage is likely to have started to have an effect (see Figure \ref{fig:squidprior}).

Without careful tuning of these parameters, we do see the data selection method fitting primarily to the yield region rather than the undamaged linear region for some of the experiments, which is unwanted. The heterogeneity of the onset of damage does make this a challenging data selection problem. However, the prior we arrive at aligns with our prior beliefs, in that we know that it is extremely unlikely for the tendon to have become damaged (and therefore not align with our model) at low strains.

\subsubsection{MCMC to infer fidelity parameters} \label{sec:MCMC-fid}

We implement a Metropolis-within-Gibbs framework to sample from the data selection posterior (Equation \eqref{eq:posterior}), alternating between updating the microstructural parameters and the fidelity parameters. The random walk step-sizes are tuned for the model and fidelity parameters independently to achieve 23.4$\pm$5\% acceptance \cite{gelman1997}. Three chains were initialised randomly for each dataset, with a burn-in of 2.5$\times10^6$ iterations and 5$\times10^6$ iterations post-burn-in. 

In order to handle the constraints on the parameters we first transformed to an unbounded scale, giving
\begin{align}
    &\gv{\chi}_\gamma = \mathcal{T}^{-1}_\gamma(\gv{\gamma}) = \logit \left(\gv{\gamma} \right) \, \in \mathbb{R}^N, \\
    &\gv{\xi} = \mathcal{T}^{-1}_{\theta}(\gv{\theta}) \, \in \mathbb{R}^4 \notag \,. 
\end{align}
For full details of $\mathcal{T}_{\theta}$ see Appendix \ref{apdx:param_transform}. The posterior in the transformed space is
\begin{equation}
    \tilde{\pi}(\gv{\xi}, \gv{\chi}_\gamma \,|\, \v{y}) \propto \tilde{\pi}_0(\gv{\xi}) \; \tilde{\pi}_0(\gv{\chi}_\gamma) \; \mathcal{L}\left(\v{y} \,|\, \mathcal{T}_\theta(\gv{\xi}),  \sigma\obs^2, 
    \mathcal{T}_\gamma(\gv{\chi}_\gamma)\right),
\end{equation}
where 
\begin{subequations}
    \begin{align}
    &\tilde{\pi}_0(\gv{\xi}) = \mathcal{N}(\gv{\xi};\gv{\mu}, \gv{\sigma}\obs^2),\\
    &\tilde{\pi}_0(\gv{\chi}_\gamma) = \mathcal{N}(\gv{\chi}_\gamma;\gv{\mu}_\gamma, \gv{\Sigma}_\gamma),\\
    \end{align}
\end{subequations}
and $\mathcal{L}\left(\v{y} \,|\, \mathcal{T}_\theta(\gv{\xi}),  \sigma\obs^2, \mathcal{T}_\gamma(\gv{\chi}_\gamma)\right)$ is as defined in Equation \eqref{eq:fidlikelihood}. 

We employed the Random walk Metropolis proposal within the Metropolis-within-Gibbs method to sample from the posterior distribution, with proposals made separately for the model parameters, and the fidelity parameters. The covariance of the microstructural parameters was learned during sampling with the proposal covariance adjusted accordingly in an adaptive manner. The covariance matrix of the proposal distributions for the fidelity parameters was chosen to be proportional to the prior covariance matrix.

\subsection{Data Truncation}
\label{sctn:data_truncation}
Following successful characterisation of the posterior distributions of the fidelity parameters, we need to truncate the data. The reason for this is twofold. Firstly, it is required as there are many observations with extremely small fidelity parameters whose effect on the posterior is negligible, but whose effect on the cost of computing the posterior is significant. Secondly, as has been seen in previous work, including data on the boundaries of regions with low posterior fidelity can lead to significant bias in the parameter estimates \cite{cotter2022hierarchical}. As such, we take the same approach as in \cite{cotter2022hierarchical}, trimming all data from the first observation with posterior fidelity mean below 0.3. Using a threshold below 0.3 leads to significant shifts in the peaks of the parameter distributions, indicating bias caused by model-data discrepancies. Increasing the threshold above 0.3 leads to the removal of data in regions where the data-model fit is still very good, lowering the information gain and increasing uncertainties.

\subsection{Inferring Population-Level Natural Variation Via a Mixed Effects Model}
\label{sctn:me_model}
Figure \ref{fig:joint_data_plot} clearly demonstrates that the values of key constitutive parameters vary between individuals. To account for and quantify this data heterogeneity, we assume that the $i$-th tendon has its own individual parameters $\gv{\theta}_i \in \mathbb{R}^2_{\geq 0} \times \mathbb{R}^2_{\geq 1}$ for $i\in\{1,2,\ldots,N_e\}$ where $N_e$ is the number of individuals within the population. In our case, $N_e=18$ for each of the two populations, SDFT and CDET, which we model separately. We denote the truncated observed stresses of the $i$-th individual as $\hat{\v{y}}_i \in \mathbb{R}^{N_i}$, and $\hat{\gv{\lambda}}_i \in \mathbb{R}^{N_i}$ as the set of stretches for which $\hat{\v{y}}_i$ was measured. The associated fidelity parameters are denoted $\gv{\gamma}_i\in [0.3,\,1)^{N_i}$. In the following, it is assumed that the data has been pre-processed and trimmed using the fidelity threshold of $0.3$ such that $N_i$ represents the length of the data after trimming. The inferred fidelity parameters are given in Section \ref{sctn:data_selection_results}.

Letting $\hat{\lambda}_{ij}$ be the $j$-th value of the stretch $\hat{\gv{\lambda}}_i$ corresponding to the $j$-th measured data point of the $i$-th experiment $\hat{y}_{ij}$, along with the $j$-th fidelity parameter $\gamma_{ij}$, then the likelihood for the $i$-th experiment can be expressed as
\begin{equation}
    \mathcal{L}(\hat{\v{y}}_i \,|\, \gv{\theta}_i,\,\sigma\obs;\, \hat{\gv{\lambda}}_i,\,\gv{\gamma}_i) \propto \prod_{j=1}^{N_i}  \frac{1}{\sigma\obs} \exp\left\{-\frac{\gamma_{ij}}{2 \sigma^2\obs}\left(\hat{y}_{ij} - \mathcal{M}(\hat{\lambda}_{ij},\, \gv{\theta}_i)\right)^2\right\}.
    \label{eq:observational_model}
\end{equation}
Experiments on different tendons are conditionally independent given the tendon-specific parameters; therefore, the joint likelihood of all experiments is the product of the likelihoods for each individual tendon. Going forward, the value of $\sigma\obs$ is chosen to be fixed with a value of 0.15 MPa$^2$, and shall be dropped from future notation.

In order to facilitate the specification of the random effects distribution, we transform the tendon-specific parameters to an unbounded scale using the same reparametrization as in Section \ref{sec:MCMC-fid}, giving
\begin{equation}
    \begin{pmatrix}
        \phi E_i\\
        (1 - \phi)\mu_i\\
        a_i\\
        b_i
    \end{pmatrix} = \gv{\theta}_i = \mathcal{T}_\theta(\gv{\xi}_i) \iff 
    \begin{pmatrix}
        \nu_i\\
        \eta_i\\
        \tau_i\\
        \rho_i
    \end{pmatrix} = \gv{\xi}_i = \mathcal{T}^{-1}_\theta(\gv{\theta}_i).
    \label{eq:invertible_transform}
\end{equation}
We assume that the unbounded parameters are samples from a common population, and model the population with a normal distribution with mean $\gv{\mu}\popu$ and covariance $\mat{\Sigma}\popu$, which is a common choice for the population distribution \cite{Betancourt_2020,LEEMEReview,Gelman_Hill_2006}. The parameters $\gv{\xi}_i$ are assumed to be conditionally independent and are distributed as $\gv{\xi}_i \,|\, \gv{\mu}\popu,\,\mat{\Sigma}\popu \sim \mathcal{N}(\gv{\mu}\popu,\,\mat{\Sigma}\popu)$. We define $\pi\popu(\cdot \,|\, \gv{\mu},\, \mat{\Sigma})$ to be the probability density function (PDF) of a normal distribution with parameters $\gv{\mu}$ and $\mat{\Sigma}$ such that $\pi\popu(\gv{\xi}_i \,|\, \gv{\mu}\popu,\,\mat{\Sigma}\popu)$ is the PDF of the population distribution.

The joint prior density of the unconstrained parameters $\pi(\gv{\xi}_1,\ldots,\gv{\xi}_{{N_e}})$ can be expressed \cite{deFinetti1972probability} as the conditionally-independent, continuous mixture
\begin{equation*}
    \pi(\gv{\xi}_1,\ldots,\gv{\xi}_{{N_e}}) = \int \pi^{\mu,\Sigma}_0(\gv{\mu}\popu,\,\mat{\Sigma}\popu) \prod_{i=1}^{{N_e}} \pi\popu(\gv{\xi}_i \,|\, \gv{\mu}\popu,\,\mat{\Sigma}\popu) \, d\gv{\mu}\popu\, d\mat{\Sigma}\popu.
\end{equation*}

For conciseness, we denote the list of parameters $\gv{\xi}_1,\ldots,\gv{\xi}_{{N_e}}$ as $\{\gv{\xi}_i\}_{i=1}^{{N_e}}$, and use similar notation for other parameters.

In a Bayesian context, we can simultaneously infer the per-individual parameters and the population parameters and then marginalise out the population parameters from the distribution. We can write the joint prior distribution of the per-individual parameters and population parameters as 
\begin{equation*}
\pi(\{\gv{\xi}_i\}_{i=1}^{{N_e}},\,\gv{\mu}\popu,\,\mat{\Sigma}\popu) = \pi^{\mu,\Sigma}_0(\gv{\mu}\popu,\,\mat{\Sigma}\popu) \prod_{i=1}^{{N_e}} \pi\popu(\gv{\xi}_i \,|\, \gv{\mu}\popu,\,\mat{\Sigma}\popu).
\end{equation*}

Finally, we must set priors for the population parameters to complete the statistical model. These are discussed in Section \ref{sctn:prior_selection}, and, for now, will be denoted $\pi^{\mu,\Sigma}_0(\gv{\mu}\popu,\,\mat{\Sigma}\popu)$. The posterior distribution can be constructed, therefore, from three parts: 1) the per-individual likelihood, 2) the population model, 3) the priors \cite{LEEMEReview}. In terms of the unconstrained parameters, the posterior distribution is given by
\begin{align}
    \begin{aligned}
        \pi(\{\gv{\xi}_i\}_{i=1}^{{N_e}},\,\gv{\mu}\popu,\,\mat{\Sigma}\popu \,|\, \{\hat{\v{y}}_i\}_{i=1}^{{N_e}},\, \{\gv{\gamma}_i\}_{i=1}^{{N_e}}, \sigma\obs ) \propto & \prod_{i=1}^{{N_e}} \mathcal{L}(\hat{\v{y}}_i\,|\,\mathcal{T}_\theta(\gv{\xi}_i),\, \hat{\gv{\lambda}}_i,\,\gv{\gamma}_i, \sigma\obs)\\
        &\cdot \prod_{i=1}^{{N_e}} \pi\popu(\gv{\xi}_i \,|\, \gv{\mu}\popu,\,\mat{\Sigma}\popu) \cdot \pi^{\mu,\Sigma}_0(\gv{\mu}\popu,\,\mat{\Sigma}\popu).
    \end{aligned}
    \label{eq:transformed_me_posterior}
\end{align}

To obtain the posterior distribution in terms of the constrained, untransformed parameters $\gv{\theta}_i$, the parameters $\gv{\xi}_i$ are transformed back to the constrained scale, which requires a Jacobian adjustment of the posterior distribution in Equation \eqref{eq:transformed_me_posterior} due to the transform $\mathcal{T}_\theta(\cdot)$, resulting in the posterior distribution of the untransformed parameters:
\begin{align}
    \begin{aligned}
        \pi(\{\gv{\theta}_i\}_{i=1}^{{N_e}},\,\gv{\mu}\popu,\,\mat{\Sigma}\popu \,|\, \{\hat{\v{y}}_i\}_{i=1}^{{N_e}},\, \{\gv{\gamma}_i\}_{i=1}^{{N_e}}, \sigma\obs) \propto & \prod_{i=1}^{{N_e}} \mathcal{L}(\hat{\v{y}}_i\,|\,\gv{\theta}_i,\, \hat{\gv{\lambda}}_i,\,\gv{\gamma}_i,\sigma\obs)\\
        &\cdot \prod_{i=1}^{{N_e}} \pi\popu(\mathcal{T}^{-1}_\theta(\gv{\theta}_i) \,|\, \gv{\mu}\popu,\,\mat{\Sigma}\popu) \abs{\d{\mathcal{T}^{-1}_\theta(\gv{\theta}_i)}{\gv{\theta}_i}}\\ &\cdot \pi^{\mu,\Sigma}_0(\gv{\mu}\popu,\,\mat{\Sigma}\popu).
    \end{aligned}
    \label{eq:untransformed_me_posterior}
\end{align}

In order to understand the population variability in the mechanical parameters, we will compute the posterior predictive distribution of the parameters $\gv{\theta}^\ast$ for a future randomly-selected tendon from the same population. This takes into account both population variability of the parameters and our uncertainty about that variability due to limited data.

\subsection{Mixed Effects Model Prior Selection}
\label{sctn:prior_selection}
Eliciting priors for use in a Bayesian context is often a non-trivial task requiring expert knowledge of the domain to inform the choice of distributions and the parameters of those distributions \cite{Stefan2022,mikkola2023,Gelman2017}. In order to evaluate the posterior density \eqref{eq:transformed_me_posterior}, we must first set priors for the parameters $\gv{\mu}\popu$, and $\mat{\Sigma}\popu$. Further, following common practices \cite{Betancourt_2020,Gelman_Hill_2006}, we do not infer the covariance matrix directly, but rather a scale matrix, $\mat{S}$, and correlation matrix, $\mat{C}$, such that $\mat{\Sigma}\popu$ can be written as
\begin{equation*}
    \mat{\Sigma}\popu = \mat{S} \mat{C} \mat{S}.
\end{equation*}
The scale matrix, $\mat{S}$, is a diagonal matrix containing the population standard deviations for each component of $\gv{\xi}_i$. The Cholesky factorisation of $\mat{\Sigma}\popu$ is therefore $\mat{S} \mat{L}_C$, where $\mat{L}_C$ is the Cholesky factor of the correlation matrix, $\mat{C}$.

Given the posterior distribution in Equation \eqref{eq:untransformed_me_posterior} and the above discussion, we must choose the prior distribution $\pi^{\mu,\,S,\,C}_0(\gv{\mu}\popu,\, \mat{S},\, \mat{C})$. We assume that the population parameters $\gv{\mu}\popu$, $\mat{S}$, and $\mat{C}$ are independent, and, therefore, the prior distribution decomposes into the product of the priors of each parameter $\pi^\mu_0(\gv{\mu}\popu)\pi^S_0(\mat{S})\pi^C_0(\mat{C})$.

In previous work \cite{haughton}, priors were set on the model parameters, $\gv{\theta}$, using distributional parameters informed by values found in the literature. Due to the hierarchical model used in this paper, we are unable to set priors directly on the $\gv{\theta}_i$ as these parameters are governed by the population distribution. Instead, priors that were previously set on the model parameters are now placed on $\gv{\mu}\popu = [\nu\popu,\,\eta\popu,\,\tau\popu,\,\rho\popu]^T$. Following the previous work, a transformation is used based on the natural bounds of the parameters, as described in Appendix \ref{apdx:param_transform}.

Conveniently, the transformation used in the previous work, which was used to boost computational efficiency by matching the support of the parameters and the MCMC transition kernel, is also an invertible transformation, as given in Equation \eqref{eq:invertible_transform}, which is required for the mixed effects model.

We set the following independent priors on the entries of $\gv{\mu}\popu$:
\begin{align*}
        \nu\popu &\sim \mathcal{N}(1.05309738,\,  1.30927056^2)\,\\
        \eta\popu &\sim \mathcal{N}(6.83672018,\,  0.47191773^2)\,\\
        \tau\popu &\sim \mathcal{N}(-3.80045123,\,  0.64387023^2)\,\\
        \rho\popu &\sim \mathcal{N}(-3.59771868,\,  0.7310165^2)\,.
\end{align*}

The prior distributional parameters were derived using values from Silver \cite{silver2002}, Purslow \cite{purslow2009}, and Goh \cite{goh2008}. We reduced the variance of the prior distributions in comparison to the previous study \cite{haughton} due to regions of high probability density of the priors coinciding with relatively unphysical parameter ranges of the model parameters in that study.

Setting priors for a correlation matrix is non-trivial. Many distributions have been proposed as priors for correlation matrices \cite{Gelman2017}, such as the Jeffreys prior, per-entry uniform priors, and a per-entry Gaussian process prior \cite{Roininen2011}. An increasingly common choice for correlation priors is the Lewandowski-Kurowicka-Joe (LKJ) distribution \cite{gelman2013bayesian,LEWANDOWSKI20091989} with parameter $\kappa > 0$. We set an $LKJ(1)$ prior on the correlation matrix, which is the default recommendation of the brms R package \cite{brms}. Finally, for the entries of the scale matrix, $S$, we set half Student t priors with 3 degrees of freedom, mean zero, and unit scale, in line with the brms defaults \cite{brms}.

Since our data appears to be very informative about the parameters in our model, the posterior is not overly sensitive to the choice of priors used here. The prior was chosen to cover values from the literature, including from collagen from skin rather than tendons, and to be overdispersive, so as to cover all possibilities and prevent an overinformative prior.

\subsection{Posterior Sampling Implementation}
\label{sctn:sampling_with_stan}
In order to characterise the mixed effects posterior distribution \eqref{eq:untransformed_me_posterior}, we use the Stan library \cite{Stan}. Below, we will discuss the specific configuration of Stan used for the inference. The sampling algorithm used in Stan is based on the HMC algorithm \cite{DUANE1987216} and its extension, the No-U-Turn sampler (NUTS) \cite{hoffman2011}.

In classical HMC sampling, the proposal vector $\gv{\theta}'$ obtained by integrating Hamilton's equations is conditionally accepted using the Metropolis acceptance ratio. In comparison, the HMC+NUTS algorithm builds a binary tree along the integrated Hamiltonian by integrating forwards and backwards with respect to the \quotes{time} variable. Rather than using a Metropolis acceptance step, the proposed state $\gv{\theta}'$ is taken to the be the point along the integrated trajectory that is furthest away from the initial state where the integration began, where integrating the Hamiltonian any further would cause the path to "turn back" on itself \cite{hoffman2011}. The number of points generated along the trajectory is controlled by a parameter $j_\mathrm{max} \geq 1$ that limits the maximum depth of the tree generated to be no more than $2^{j_\mathrm{max}} - 1$. In Stan, this parameter is called \texttt{max\_tree\_depth} and defaults to 10.

Stan automatically tunes the leapfrog parameters and the Euclidean metric during an adaptive warmup period before sampling. More information about these parameters, and what they mean in the context of the HMC+NUTS algorithm can be found in the Stan user manual \cite{Stan}. The user is able to control the adaptation of the integrator parameters by changing the values of user-facing parameters called \texttt{step\_size} and \texttt{adapt\_delta}. The \texttt{step\_size} parameter defines the starting guess for the integration step size before the adaptation phase. The parameter \texttt{adapt\_delta} acts as a surrogate for a target Metropolis acceptance ratio, and controls the adaptation of the number of integration steps, the estimation of the Euclidean metric, and the integration step size.

If the HMC+NUTS parameters are not set properly, or are unable to adapt to optimal values, then the integrator will accrue enough errors that the integrated Hamiltonian trajectory will not follow the true trajectory. This divergence from the true trajectory is reported to the user, and divergences indicate poor exploration of the state space and the simulation cannot be trusted. With the posterior distribution we are sampling from in this study, the default parameter values resulted in divergences, poor effective sample sizes, and non-convergence to the posterior as reflected by the split R-hat statistic. Instead, we used the following values for the parameters: \texttt{step\_size} = 0.01, \texttt{adapt\_delta} = 0.99, and \texttt{max\_tree\_depth} = 14. To interface with Stan, we used cmdstanr \cite{cmdstanR} with a modified Stan script generated from brms \cite{brms}.

For our analysis, we generated 4000 samples per chain, with 10 chains running in parallel, for a total of 40,000 posterior samples. The average sampling time per chain was around 60 hours between both runs using a server with an Intel(R) Xeon(R) CPU E5-4627 v2 @ 3.30GHz (32 core) with 64GB memory, with each chain utilizing one core.

\section{Results}
\label{sctn:results}We provide numerical results for the data selection method, introduced in Section \ref{sctn:fidel_params}, in Section \ref{sctn:data_selection_results}. Using these results, we truncate the data as explained in Section \ref{sctn:data_truncation}, and feed the truncated data into the mixed effects inference. Results for the mixed effects model, using the data fed from the data selection method, are given in Section \ref{sctn:me_model_results}.

\subsection{Data Selection Results}
\label{sctn:data_selection_results}

\begin{figure}[h]
    \centering
    \includegraphics[width=\linewidth]{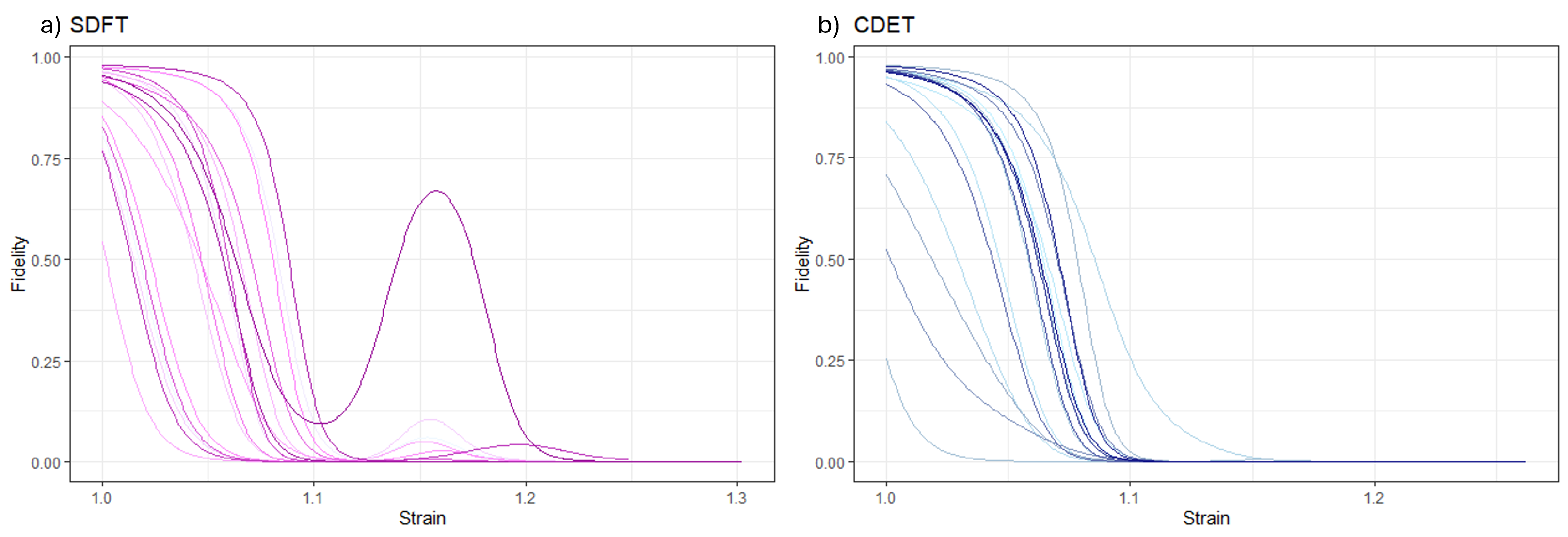}
    \caption{Posterior means of the fidelity parameters against strain for a) SDFT and b) CDET tendon samples. Each line represents a fidelity profile for a single tendon.}
    \label{fig:all-fid}
\end{figure}

\begin{figure}[h]
    \centering
    \includegraphics[width=\linewidth]{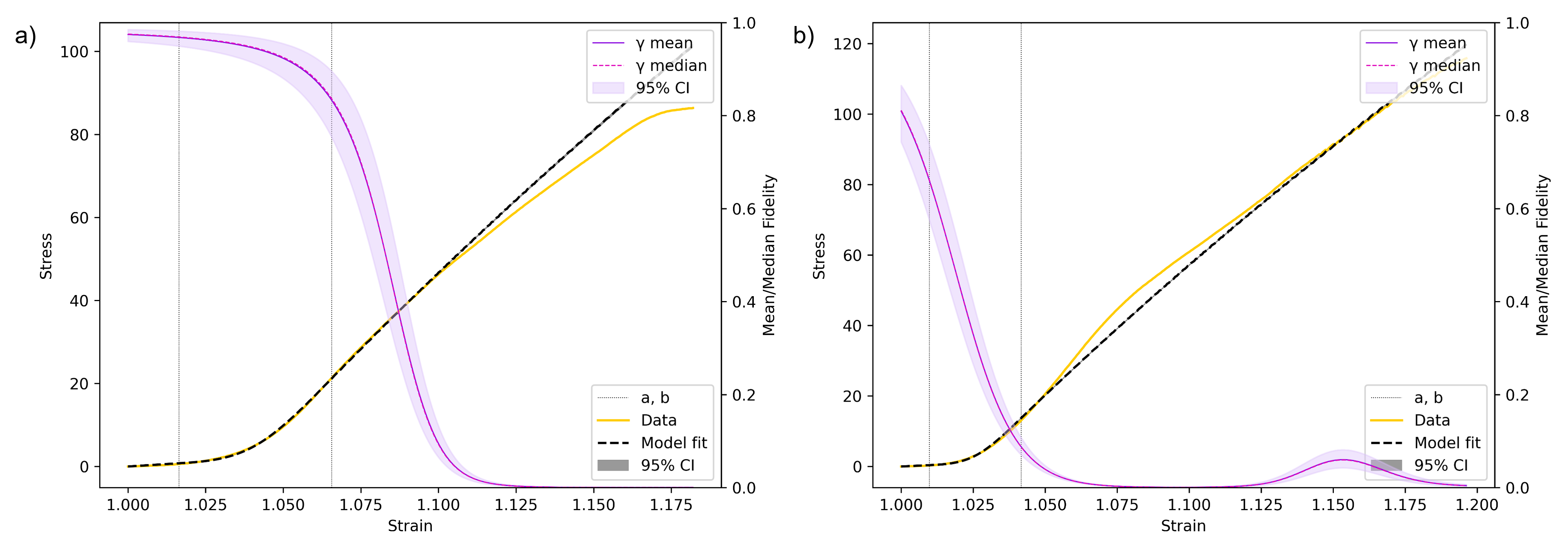}
    \caption{Comparison of the model fit for two SDFTs and the fidelity parameter profiles. a) A tendon (H16 SDFT) where the fidelity parameters decrease with strain, and b) a tendon (H15 SDFT) where the fidelity parameters decrease but then increase again at a larger strain due to model-data crossing. }
    \label{fig:model-fid-comp}
\end{figure}

Each individual tendon sample dataset was fit using the microstructural model with data selection. Stress-strain data was extracted up to the point of maximum stress as the model is elastic and, therefore, not capable of describing subsequent decreases in stress that are associated with damage. The model and fidelity parameters were inferred using the MCMC methodology outlined in Section \ref{sec:MCMC-fid}. Here, we focus on the fidelity parameter marginal means, which are subsequently used within the mixed effects model. The majority of samples exhibited fidelity parameter profiles which started at values close to one, and decreased to zero at some point between 5\% and 10\% strain (see Figures \ref{fig:all-fid} and \ref{fig:model-fid-comp}a). Some fidelity profiles, however, exhibited a secondary, smaller increase beyond 10\% strain. This was simply due to the model and data crossing at larger strains. It was evident that the model had already diverged from the data by this point and the increase was simply an artefact of model extrapolation (see Figure \ref{fig:model-fid-comp}b).

\begin{figure}[h]
    \centering
    \includegraphics[width=\linewidth]{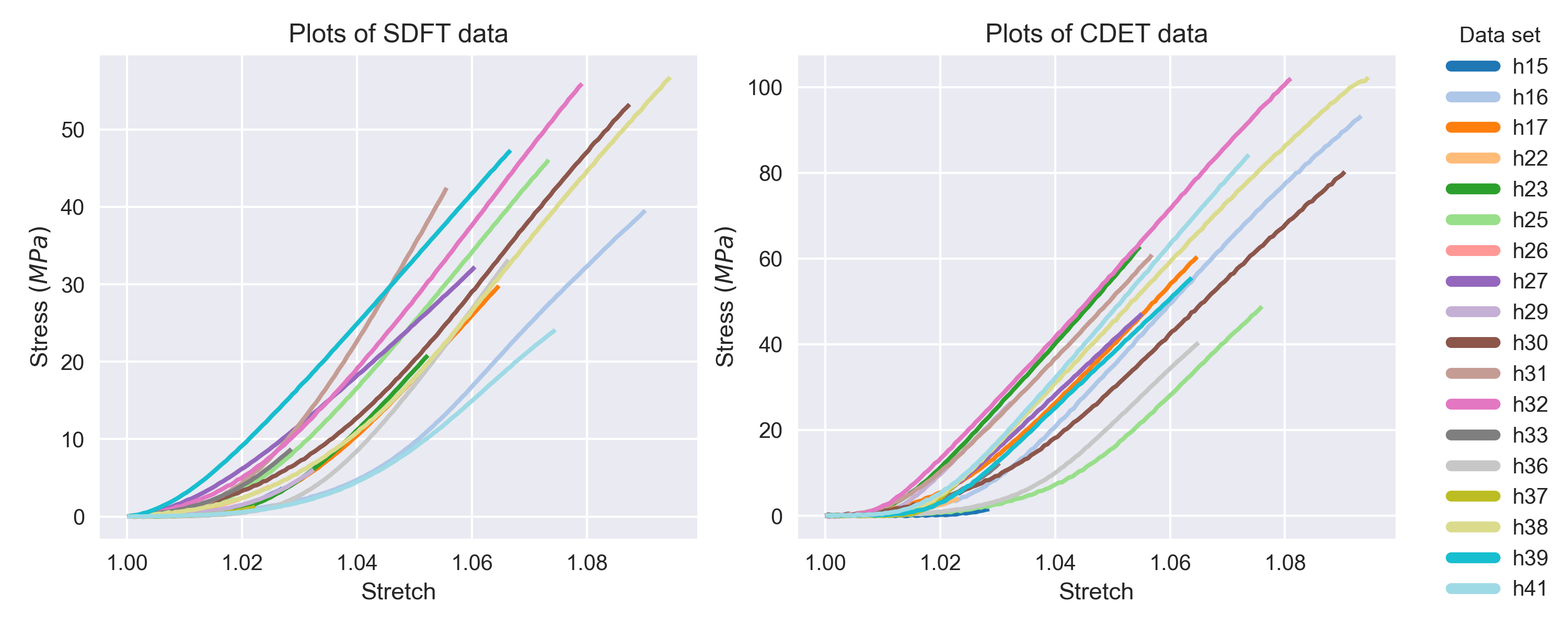}
    \caption{Plots of experimental tensile test data for SDFT (left) and CDET (right), trimmed automatically using Bayesian data selection and a fidelity threshold of 0.3.}
    \label{fig:joint_data_plot_trimmed}
\end{figure}

Interestingly, the fidelity parameters of all the tendons reduced to values below 0.3 before 10\% strain, indicating a significant reduction in the contribution of the data to the likelihood beyond this point. This is shown in Figure \ref{fig:joint_data_plot_trimmed} where the stress-strain profiles of the SDFTs and CDETs have been trimmed to the point where the fidelity parameters first go below 0.3. There is considerable variability in the strain at which the fidelity parameters first reduce below 0.3, in both the SDFTs and CDETs. This further supports our approach of using data selection and the inference of fidelity parameters as opposed to homogeneous manual trimming of the data. 

\subsection{Mixed Effects Model Results}
\label{sctn:me_model_results} 
Next, we use the truncated data and fidelity parameter posterior means to infer the population-level distributions of the constitutive parameters. We used the MCMC methods described in Section \ref{sctn:sampling_with_stan} to sample from the posterior distribution of the mixed effects model as outlined in Section \ref{sctn:me_model}. The mechanical model consists of two elastic parameters, $(1-\phi)\mu$ and $\phi E$, and two structural parameters, $a$ and $b$. Due to the nature of the transformation $\mathcal{T}_\theta(\cdot)$ and the priors, the  posterior predictives for the parameters are best understood through their modes and not their means. This is because the posterior distributions are heavy tailed, and may not necessarily have finite means or variances.

\subsubsection{Posterior predictives for model parameters}
\begin{figure}[!ht]
    \begin{subfigure}[t]{0.48\textwidth}
        \centering
        \includegraphics[width=\linewidth]{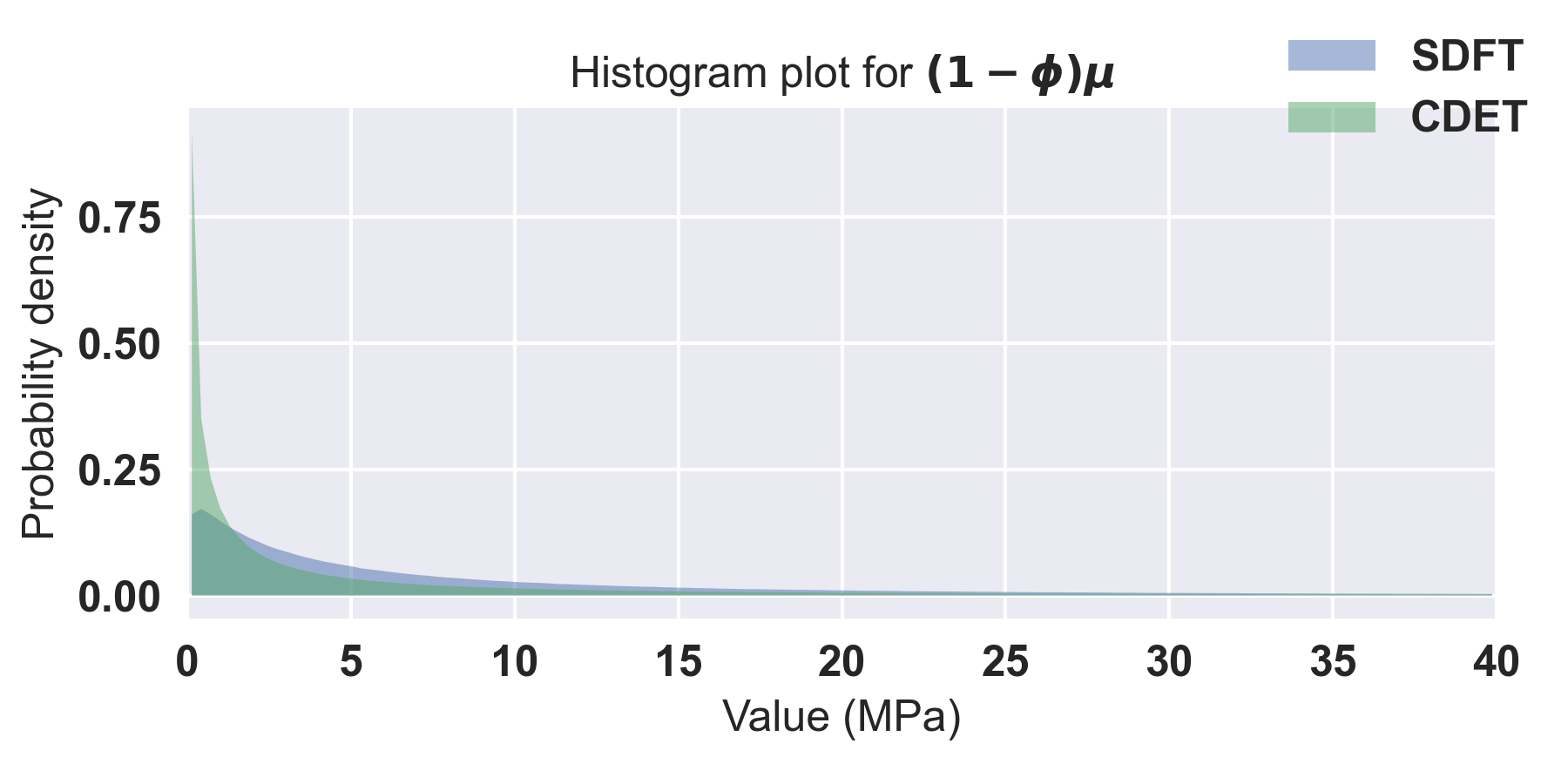}
    \end{subfigure}
    \hfill
    \begin{subfigure}[t]{0.48\textwidth}
        \centering
        \includegraphics[width=\linewidth]{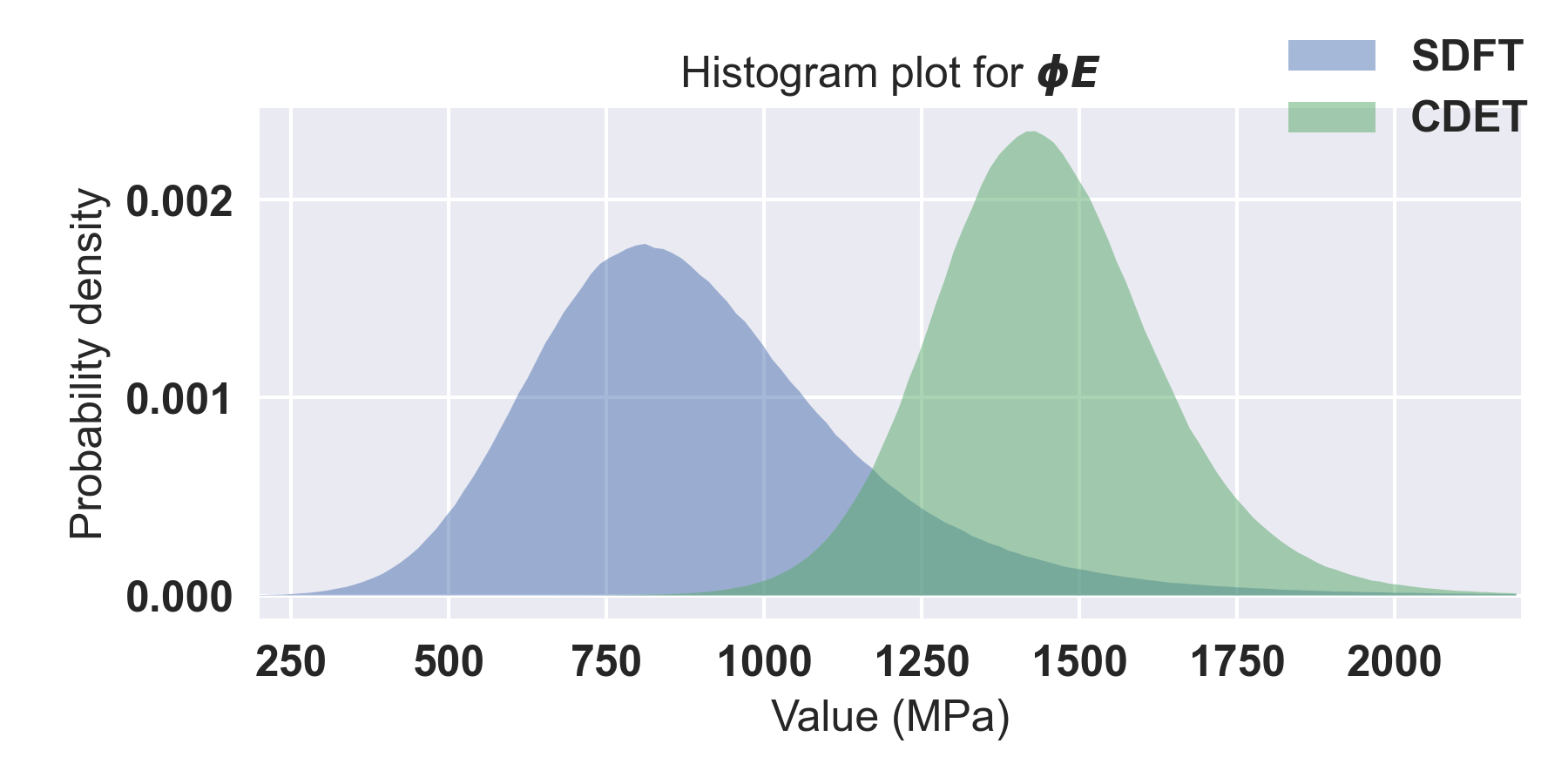}
    \end{subfigure}
    \vfill
    \begin{subfigure}[t]{0.48\textwidth}
        \centering
        \includegraphics[width=\linewidth]{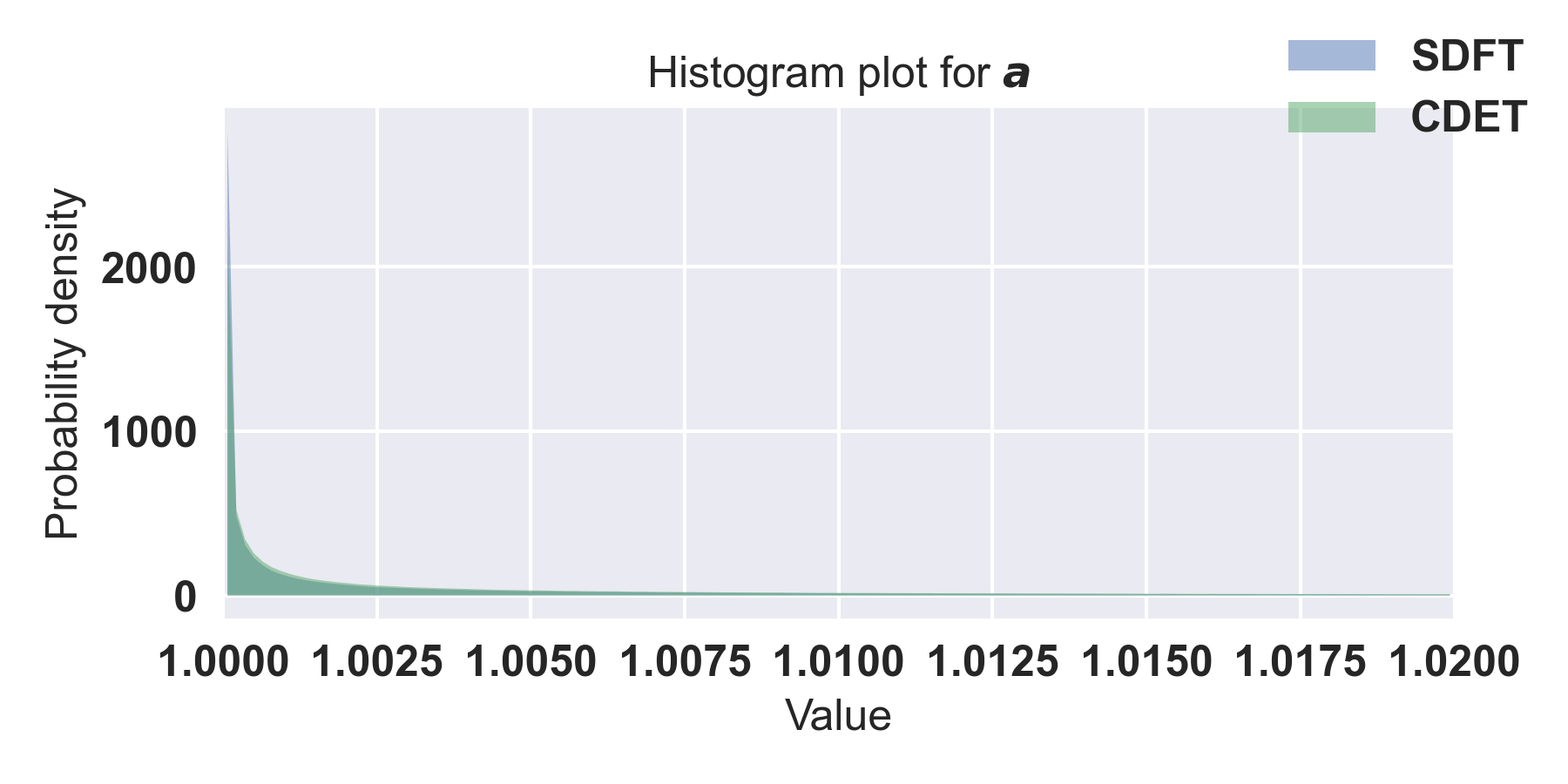}
    \end{subfigure}
    \hfill
    \begin{subfigure}[t]{0.48\textwidth}
        \centering
        \includegraphics[width=\linewidth]{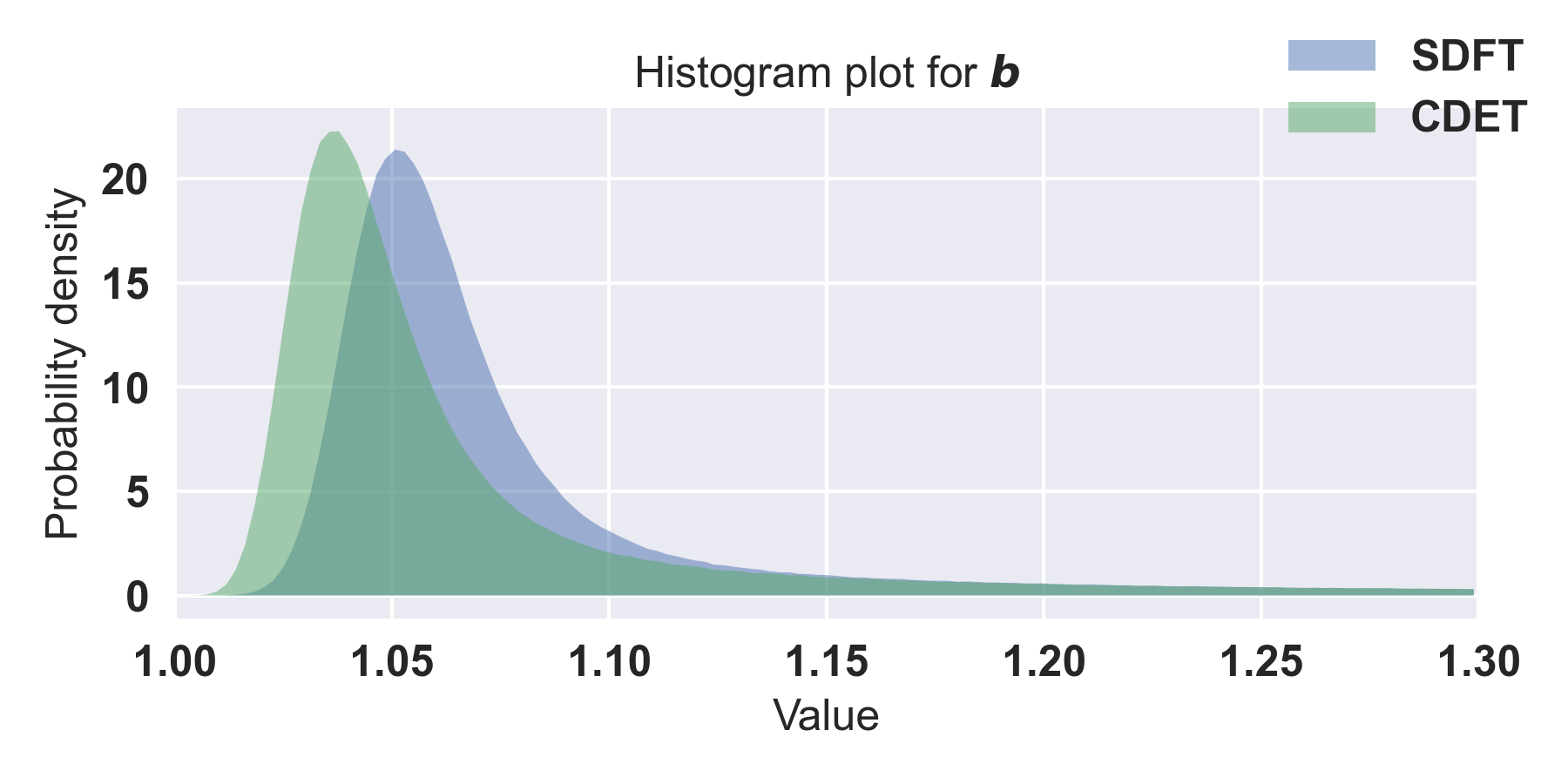}
    \end{subfigure}
    \caption{Comparison of the marginal posterior predictive distribution for each model parameter.}
    \label{fig:comparison_plots}
\end{figure}

The posterior predictives of the model parameters are compared for the SDFTs and CDETs in Figure \ref{fig:comparison_plots}. For the CDET distributions, $\phi E$ has a peak at 1430.2 MPa, whereas the corresponding peak for the SDFT lies at 811.5 MPa. Additionally, the CDET population distribution for $(1-\phi)\mu$ has density closer to 0 in comparison to the SDFT population distribution. Both of these quantities include the non-identifiable structural parameter $\phi$. There are two possible causes, which in some combination could give rise to the differences that we see in the mechanical parameter distributions. The CDET could contain fibres that are stiffer than those found in the SDFT, and/or the CDET could contain a larger volume fraction of fibrils than the SDFT.

For the structural parameters, $a$ and $b$, the distributions for the SDFT tend to favour higher values than those of the CDET. The parameter, $b$, has peak probability density at a value of 1.05 for the SDFT, a larger value than the CDET peak density value of 1.04, and the spread of values is larger for the SDFT than the CDET.

These differences in peak probability density relate to the biological functions of the tendons. As the SDFT is an energy-storing tendon that extends and contracts to facilitate locomotion \cite{Thorpe2012}, the higher values for the structural parameters reflect the larger range of strains the tendon operates over and the correspondingly longer collagen fibril lengths. The lower values of the mechanical parameters indicate lower stiffness, complementing the functionality of the tendon. The higher stiffness and lower values for structural parameters of CDET reflect their relative inextensibility, which facilitates the transfer of force between muscle and bone.

It is worth noting that these distributions do not solely represent the natural variation of the parameters in the population, but also the uncertainty in these distributions due to the relatively small number of individuals that have been used in the inference. As such, it is likely that these distributions are overdispersed compared to the true underlying distributions. It is infeasible to disentangle this uncertainty from the natural variation of the parameters, but with the addition of further data, the posterior would likely contract further.

\edit{
%In their original paper, Thorpe \etal\cite{Thorpe2012} calculated structural and mechanical properties of the tendon samples. They provided estimates of the linear modulus of the SDFT and CDET by calculating the gradient every 5 data points, smoothing these estimates using a moving average of window length 5, and taking the maximum value. 

The tangent modulus in the linear region of a tendon's stress-strain curve is often reported as an important mechanical quantity \cite{Thorpe2012}. As a naive estimate, the linear modulus might be used as an approximation of $\phi E$; however, as we showed in Section \ref{sctn:tendon_model}\ref{sctn:linear_modulus}, an exact calculation of the linear modulus of our model shows that it will be strictly less than $\phi E$. To explore whether the linear modulus can be used as a reasonable estimate of $\phi E$, we compare our posterior distributions to distributions of the linear modulus of the respective tendon type. %To calculate the linear modulus, we determined the linear region using the fidelity threshold and the inferred values of $b$ as the upper and lower bounds, respectively. 
To define the linear region, for the upper bound, we used the cutoff stretch as calculated in Section \ref{sctn:data_selection_results}, and for the lower bound, the posterior mean of samples of $b$ from \emph{independent} inferences of each data set was used. For the SDFT, 12 out of 18 data sets had valid detected linear regions, where the lower bound is less than the upper bound, and 16 of the 18 data for the CDET had valid detected linear regions. From the data with valid linear regions, we used linear regression to find the gradient of a straight line fit to the data as the linear modulus. Then, motivated by $\phi E$ being strictly non-negative, and the lognormal-like priors as described in Section \ref{sctn:prior_selection} along with the parameter transform described in Appendix \ref{apdx:param_transform}, in Figure \ref{fig:phiE_comparison_to_Linear_Modulus} we plot lognormal distributions for the linear moduli with mean and variance equal to the mean and variance of the linear moduli calculated from the regression data.The means of the plotted distributions are shown as vertical lines. As anticipated, using the linear modulus as a direct estimate of $\phi E$ underestimates its value compared to the inferences from our model, however there is considerable overlap in the assumed distribution of the linear modulus and our posterior distributions for $\phi E$.

% Thorpe \etal\cite{Thorpe2012} provided a mean and a standard deviation for the linear modulus of the SDFT and CDET. These values, however, appear to underestimate the Young's modulus more than can be accounted for by a factor of $\lambda^2$. We, therefore, fitted a straight line to the linear region of each data and the gradient was taken to be the linear modulus. The values calculated were $859.63 \pm 175.82$ MPa for SDFT and $1333.16 \pm 137.47$ MPa for CDET, and it is noted that these values are larger than those reported by Thorpe \etal\cite{Thorpe2012}. These values are used as the mean and standard deviation of a normal distribution to compare with our posterior distributions for $\phi E$ in Figure \ref{fig:phiE_comparison_to_Thorpe_Linear_Modulus}. As anticipated, using the linear modulus as a direct estimate of $\phi E$ underestimates its value compared to the inferences from our model.

\begin{figure}
    \centering
    \begin{subfigure}[t]{0.48\textwidth}
        \centering
        \includegraphics[width=\linewidth]{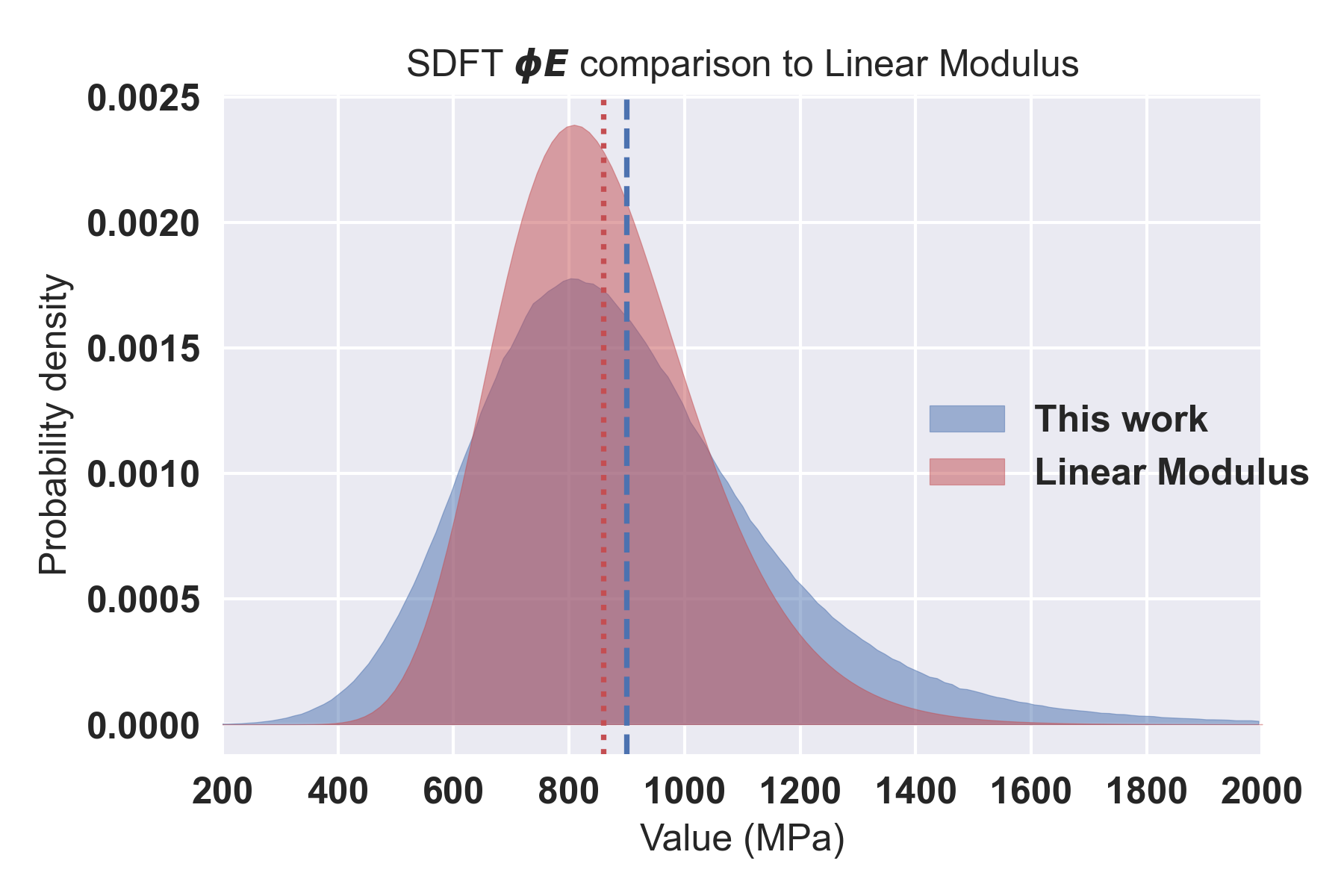}
    \end{subfigure}
    \begin{subfigure}[t]{0.48\textwidth}
        \centering
        \includegraphics[width=\linewidth]{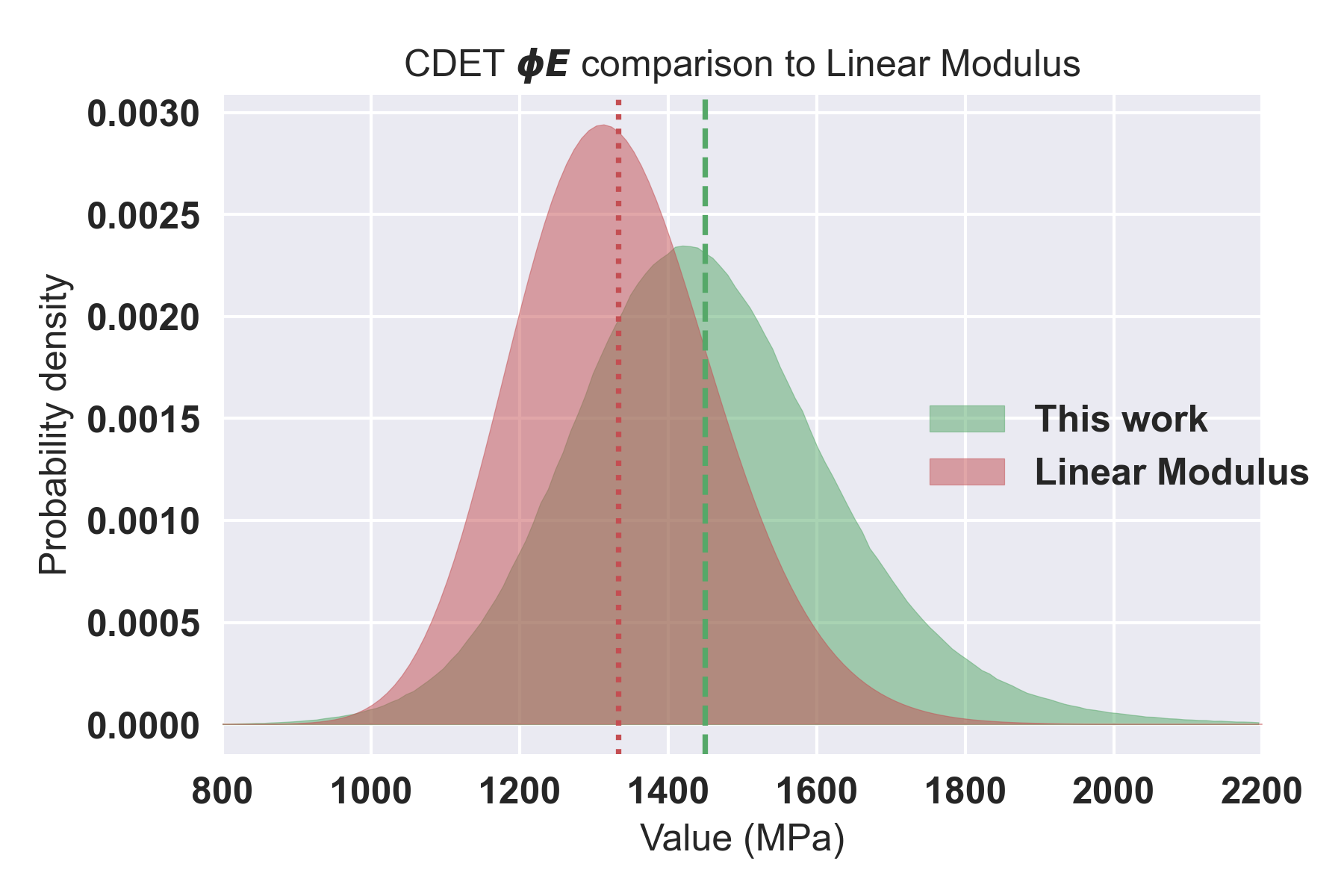}
    \end{subfigure}
    \caption{\edit{Estimates of the posterior distributions of $\phi E$ for the SDFT (left, blue) and CDET (right, green) compared to lognormal distributions with prescribed means and variances equal to the mean and variance of the linear moduli as calculated from the data (both, red). The means of each distribution are indicated with vertical lines (linear modulus: dotted; SDFT/CDET: dashed) in the corresponding colours for their respective distributions.}}
    \label{fig:phiE_comparison_to_Linear_Modulus}
\end{figure}
}

\subsubsection{Posterior predictives for the stress}
In order to visualize the fit and the amount of data included in the inference in context, we considered the posterior predictive distribution for the response of each tendon. This can be understood as the marginal posterior distribution of new stresses $\gv{y}^\ast_i$ observed in a hypothetical replication of the experiment for the $i$-th tendon at the same stretch levels as in the original data, and can be simulated by evaluating the microstructural model $\mathcal{M}(\cdot)$ at the observed stretches $\gv{\lambda}_i$, using posterior samples for the model parameters $\gv{\theta}_i$.

\begin{figure}[!ht]
    \begin{subfigure}[t]{0.48\textwidth}
        \centering
        \includegraphics[width=\linewidth]{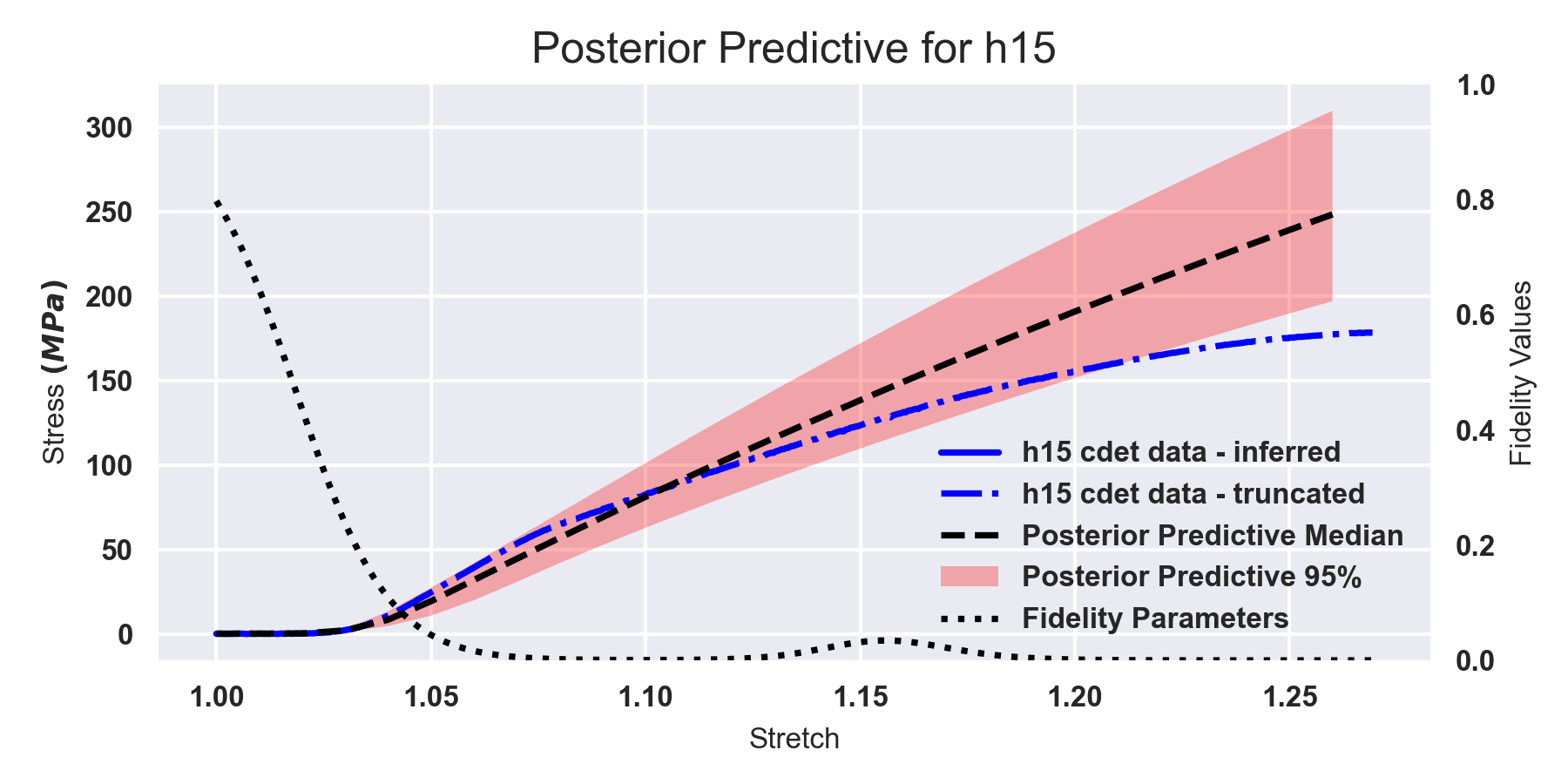}
    \end{subfigure}
    \hfill
    \begin{subfigure}[t]{0.48\textwidth}
        \centering
        \includegraphics[width=\linewidth]{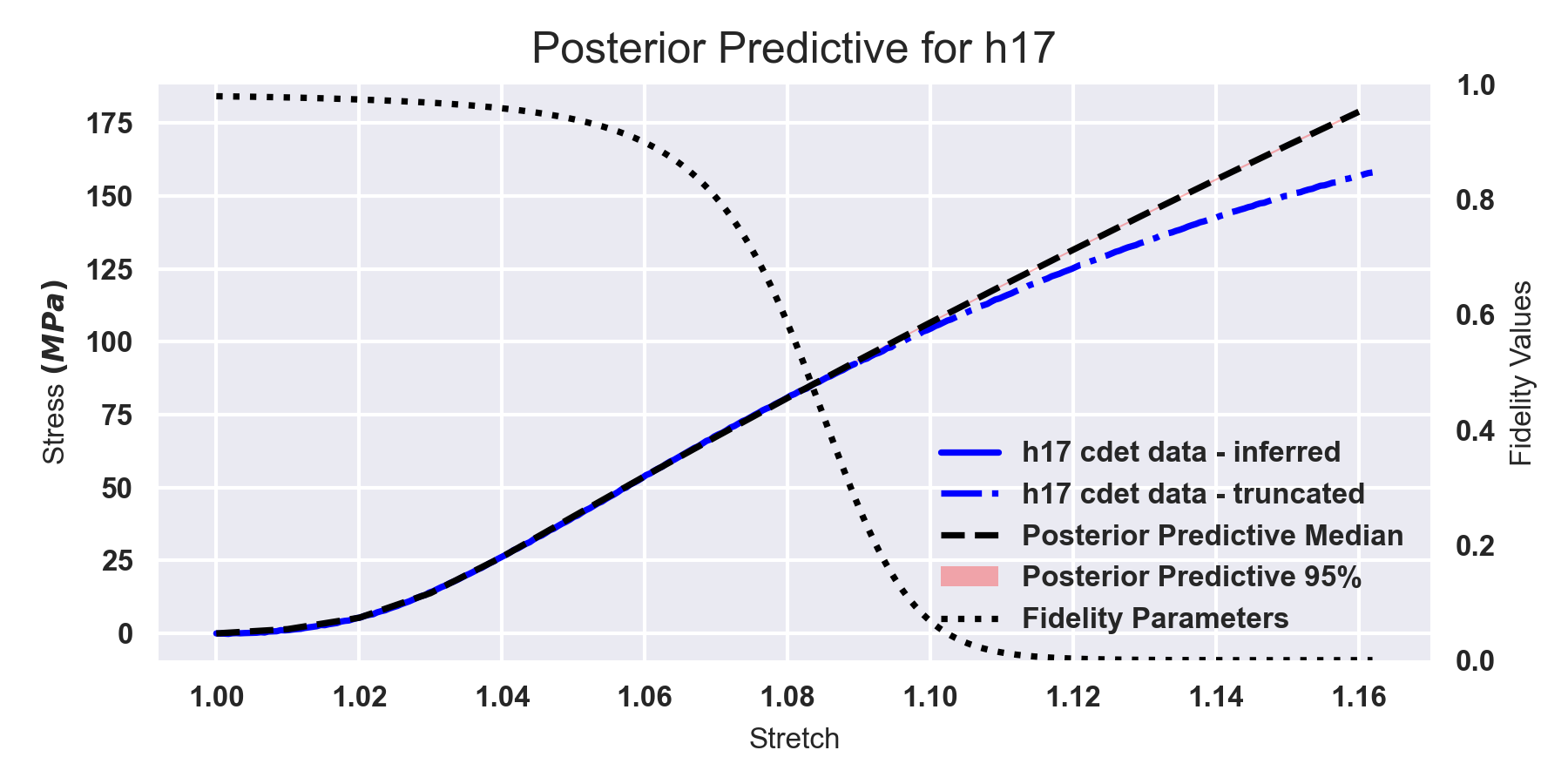}
    \end{subfigure}
    \caption{Posterior predictives for the CDET datasets h15 and h17. For h15, there are not many data points as determined by the fidelity threshold chosen, whereas h17 has many data points. The difference in inferential power is seen by the 95\% spread of the predictive. The large spread is due to the lack of information about the mechanical parameters $(1 - \phi)\mu$ and $\phi E$.}
    \label{fig:cdet_posterior_predictives}
\end{figure}

Figure \ref{fig:cdet_posterior_predictives} shows posterior predictives for the response of two CDET individuals that have differing numbers of data points after being trimmed using our data selection method. We see that, for data with fidelity parameters that do not decay until well within region III of the stress response, there is a very tight predictive with a small standard deviation about the median. In contrast, where much of the data has been removed in the selection process, the inference results in a diffuse predictive with a high standard deviation. The reduction in the amount of inferred data included in the inference greatly increases the uncertainty in the posterior predictive distributions of the parameters in experiments with small $N_i$. However the data selection methodology enabled us to fit our model to regions where the probability of the data being consistent with the chosen model was highest, ultimately giving a truer reflection of the constitutive parameter values.

\section{Discussion}
\label{sctn:discussion}In this paper, we introduced a two-stage process for producing higher quality inferences from stress-strain tests to failure, where the mechanical model that we fitted to the data does not account for damage or failure. Several studies have proposed methods to identify the start of the yield region, often relying on curve-fitting or localised gradient estimates. Techniques such as moving average smoothing of tangent moduli~\cite{Thorpe2012}, spline fitting for yield point detection~\cite{peloquin2016advances}, and polynomial fitting to identify inflection points~\cite{goh2012bimodal} offer practical means of identifying which data should be used for inference, though they rely on heuristic choices. Others have noted that there can be challenges in parameter identifiability, particularly when relying on uniaxial data alone~\cite{safa2021identifiability}.

The first stage of our approach involved inferring fidelity parameters which measure how consistent the model is with the data. By using a Gaussian process prior for the underlying fidelity field, we were able to learn the different regions of observation space for which the model is not valid. After inferring fidelity parameters independently for each data set, a fidelity threshold was chosen which was used to trim each data set automatically, and the observational noise value for each observation was scaled by its fidelity value, effectively tuning out data points which are less consistent with the model. As used in previous work \cite{cotter2022hierarchical}, we chose a threshold of 0.3. 

Using the Bayesian data selection method, we obtained higher quality inferences which accurately trimmed the data on a per-individual basis, to minimise bias, and to maximise information quality. This was necessary because of the heterogeneity of the data between experiments, leading to the model being a valid representation of the data for different strain ranges in each case. This research provides a basis for improving the predictive power of results derived from Bayesian inferences and point estimates as our data selection method is agnostic to the statistical method chosen in the second stage. It is clear from Figures \ref{fig:joint_data_plot} and \ref{fig:joint_data_plot_trimmed} that capturing the heterogeneity in damage is important in the modelling process and ignoring it will skew parameter estimates away from their true values. It is important to note the additional computational cost of our approach; an additional inference was required for each of the tendon experiments in order to appropriately trim the data. However, these can be conducted in parallel, and on modern computer infrastructures this can be achieved straightforwardly. The additional cost of the data selection is still relatively small in comparison with the cost of characterising the Bayesian mixed effects posterior, which has a large number of correlated parameters that make mixing of the Markov chain challenging.

Following the data selection stage, the trimmed and re-weighted data was fed into a Bayesian mixed effects statistical model which was used to infer population-level parameters for the mechanical model. We found that the product of the collagen volume fraction and collagen fibril Young's modulus, $\phi E$, has posterior modal values of 811.5 MPa for the SDFT, and 1430.2 MPa for the CDET. We found that CDETs are stiffer than SDFTs, due either to having stiffer fibrils, a higher collagen volume fraction, or both. We also found that the SDFTs have longer fibrils on average.

The better a model is at representing data, the more of the data can be used to infer its parameters. For example, using models which attempt to model damage or failure of the tendon \cite{JGregory,HAMEDZADEH2018483} will increase the number of data points that are valid under the assumptions of the model, and therefore less data will be lost to trimming. The use of a more computationally costly mechanical model, however, will add to the computational complexity of characterising the posterior distribution. Additionally it is very likely, however sophisticated and flexible the chosen model is, that experimental data will still require some level of selection prior to analysis in order to arrive at accurate parameter estimates. Combining sophisticated data selection methodology with mixed effects models enabled us to analyse the natural variability of parameters which define the physical properties of horse tendons more accurately than previous approaches.

\section*{Data accessibility}
The code used to implement the methods explained in Sections \ref{sctn:fidel_params} and \ref{sctn:me_model} is available at:\\
\url{https://github.com/JJ-Casey/TendonMEDS-Code}.\\
The raw data \cite{goh2008} used in this study is available at: \url{https://qmro.qmul.ac.uk/xmlui/handle/123456789/13395}

\section*{Acknowledgements}
JF was supported by the Engineering and Physical Sciences Research Council (EPSRC) [grant number EP/W522466/1].
JC was supported by the Engineering and Physical Sciences Research Council (EPSRC) [grant number EP/T517823/1] 

\bibliographystyle{abbrv}
\bibliography{thebibliography}

\begin{appendices}

\section{Parameter Transform}
\label{apdx:param_transform}
In \cite{haughton}, a transformation of the model parameters $\gv{\theta}$ was used to improve the computational efficiency of the random walk Metropolis-Hastings algorithm. The transformation arises from considering the natural bounds on the model parameters. The parameters, $E$ and $\mu$, are greater than 0, and $\phi$ is a parameter constrained to the interval $[0,\,1]$; therefore, $(1-\phi)\mu$ and $\phi E$ are are non-negative. Thus, a logarithmic transform is appropriate. The structural parameter, $a$, is greater than 1, and $b$ is greater than $a$, so we consider the parameters $a-1$ and $b-a$ which are both greater than 0 and such that their logarithms exist on the entire real line. The transform $\mathcal{T}_\theta(\cdot)$ and its inverse $\mathcal{T}^{-1}_\theta(\cdot)$ is therefore given by
\begin{equation*}
    \mathcal{T}^{-1}_\theta(\gv{\theta}) = \gv{\xi} = 
    \begin{pmatrix}
        \nu\\
        \eta\\
        \tau\\
        \rho
    \end{pmatrix}
    =
    \begin{pmatrix}
        \ln\left\{ (1 - \phi)\mu \right\}\\
        \ln\left\{ \phi E \right\}\\
        \ln\left\{ a - 1 \right\}\\
        \ln\left\{ b - a \right\}
    \end{pmatrix}
    \iff
    \gv{\theta} = \mathcal{T}_\theta(\gv{\xi}) =  
    \begin{pmatrix}
        \exp\{\nu\}\\
        \exp\{\eta\}\\
        \exp\{\tau\} + 1\\
        \exp\{\rho\} + \exp\{\tau\} + 1
    \end{pmatrix}.
\end{equation*}

\end{appendices}

\end{document}